\documentclass[manuscript,screen]{acmart}

\usepackage{algorithm}
\usepackage{algorithmic}
\usepackage{enumitem}
\usepackage{multirow}
\usepackage{pifont}
\usepackage{subcaption}
\usepackage{xspace}

\AtBeginDocument{%
  }

\setcopyright{none}
\acmJournal{TOCS}
\begin{document}

\makeatletter
\newcommand{\ifAnonymous}[2]{%
	\if@ACM@anonymous #1\else #2\fi}
\makeatother

\newcommand{\sysprefix}{Prism}
\newcommand{\sysname}{\sysprefix LLM\xspace}
\newcommand{\graphname}{\sysprefix Graph\xspace}
\newcommand{\graphsubsysname}{\textit{Graph Construction Subsystem}\xspace}
\newcommand{\emulationsubsysname}{\textit{Hybrid Emulation Subsystem}\xspace}

\newcommand{\noindentpar}[1]{\noindent \textbf{#1}}
\newcommand{\secref}[1]{\S\ref{#1}}
\newcommand*\circled[1]{\ding{\number\numexpr#1+171\relax}}
\newcommand*\smalltt[1]{\texttt{\small #1}}

\title{A Few GPUs, A Whole Lotta Scale: Faithful LLM Training Emulation with \sysname}

\author{Boyi Jia}
\orcid{0009-0001-7082-019X}
\affiliation{%
	\institution{Shanghai Jiao Tong University}
	\city{Shanghai}
	\country{China}}
\affiliation{%
	\institution{Alibaba Group}
	\city{Beijing}
	\country{China}}
\email{jiaboyi.jby@sjtu.edu.cn}

\author{Shaoke Xi}
\orcid{0009-0008-7804-3867}
\authornote{These authors contributed equally to this work.}
\affiliation{%
	\institution{Alibaba Group}
	\city{Beijing}
	\country{China}}
\email{shaoke.xsk@alibaba-inc.com}

\author{ChonLam Lao}
\orcid{0000-0002-0214-664X}
\authornotemark[1]
\affiliation{%
	\institution{Alibaba Group}
	\city{Seattle}
	\country{USA}}
\affiliation{%
	\institution{Harvard University}
	\city{Boston}
	\country{USA}}
\email{chonlamlao@g.harvard.edu}

\author{Jiaqi Gao}
\orcid{0000-0003-3548-2030}
\authornotemark[1]
\correspondingauthor
\affiliation{%
	\institution{Alibaba Group}
	\city{Seattle}
	\country{USA}}
\email{jiaqi.g@alibaba-inc.com}

\author{Chenlin Xie}
\orcid{0009-0000-7383-4004}
\affiliation{%
	\institution{Alibaba Group}
	\city{Beijing}
	\country{China}}
\email{chenlin.xcl@alibaba-inc.com}

\author{Wenting Shen}
\orcid{0000-0003-3775-6234}
\affiliation{%
	\institution{Alibaba Group}
	\city{Beijing}
	\country{China}}
\email{wenting.swt@alibaba-inc.com}

\author{Jiamin Cao}
\orcid{0009-0009-5468-7366}
\affiliation{%
	\institution{Alibaba Group}
	\city{Beijing}
	\country{China}}
\email{caojiamin.cjm@alibaba-inc.com}

\author{Erci Xu}
\orcid{0000-0002-6654-4364}
\correspondingauthor
\affiliation{%
	\institution{Shanghai Jiao Tong University}
	\city{Shanghai}
	\country{China}}
\email{jostep90@gmail.com}

\author{Ang Wang}
\orcid{0009-0007-2650-0504}
\affiliation{%
	\institution{Alibaba Group}
	\city{Beijing}
	\country{China}}
\email{wangang.wa@alibaba-inc.com}

\author{Ennan Zhai}
\orcid{0000-0003-4352-7497}
\affiliation{%
	\institution{Alibaba Group}
	\city{Hangzhou}
	\country{China}}
\email{ennan.zhai@alibaba-inc.com}

\renewcommand{\shortauthors}{B. Jia et al.}

\begin{abstract}

  Large language model (LLM) training today runs on clusters spanning thousands
  of GPUs. While this scale enables rapid model advances, developing, debugging,
  and performance-tuning the training infrastructure inevitably become complex
  and costly. This is because engineers often need to reproduce production
  behaviors or evaluate optimizations, thereby demanding frequent and even
  exclusive access to production-scale clusters---which becomes increasingly hard
  given that the majority of GPUs are already committed to production workloads.
  Simulation relies on complex performance models that are difficult to maintain,
  and downscaled experiments often fail to capture scale-dependent behaviors.

  We present \sysname, a system that decouples large-scale execution from large
  physical clusters, enabling engineers to run and observe ranks of interest with
  faithful at-scale behavior using only a few GPUs. \sysname constructs a
  high-fidelity execution representation by capturing computation, communication,
  and dependencies via context switching, while measuring and aligning timings
  across ranks via a slicing-based approach. Using this representation, \sysname
  performs hybrid emulation in which selected ranks execute the original program
  while the remaining ranks are replayed as virtual participants.

  Experiments on large-scale LLM training workloads show that \sysname accurately
  emulates performance and memory behavior, achieving only 0.58\%
    average error in iteration time, 0.65\% under straggler effects, and less than
    0.01\% error in peak GPU memory usage. \sysname can emulate clusters of up to
  8192 GPUs using fewer than 1\% of the physical GPUs required by the original
  deployment. In production, we build a training configuration tuning
    loop driven by \sysname, reducing evaluation GPU hours by 87.5\% and
    outperforming expert-tuned baselines by up to 3.10\% for online training
    tasks.

\end{abstract}




\maketitle

\section{Introduction}
\label{sec:intro}

Training large language models (LLMs) has been scaling at an unprecedented
speed. Production systems today routinely involve thousands to tens of
thousands of GPUs~\cite{llama3, deepseekv3, gpt4, qwen3, astral}, and emerging
frontier models are expected to push this scale even further. While significant
research has focused on improving training performance at such
levels~\cite{megatron-lm, deepspeed, pytorch_fsdp,fa1,fa2,fa3,fa4}, much less
attention has been paid to the \emph{cost} of developing, debugging, and
optimizing the infrastructure that supports these workloads.

In practice, LLM training follows a DevOps workflow in which engineers
continuously analyze training telemetry, implement optimizations, and deploy
updates to production jobs. This workflow critically depends on the ability to
reproduce and analyze field behavior under realistic conditions. Unfortunately,
as cluster scale grows, so does the cost of system experimentation. Reproducing
a performance issue, validating a design change, or conducting a what-if
analysis often requires access to the same production-scale cluster that runs
the training job~\cite{robust_bytedance, straggler_bytedance}. Allocating
thousands of GPUs solely for debugging or experimentation is both expensive and
operationally disruptive, making rapid system iteration increasingly difficult.

Two existing solutions partially alleviate the issue, but neither suffices. One
option is
simulation~\cite{astrasim1,astrasim2,simai,vtrain,multiverse,daydream,dpro,distsim},
which lowers hardware cost by replacing real execution with profiled traces and
analytical models. But simulation accuracy depends on the fidelity of those
models, and maintaining them is increasingly difficult as training stacks
evolve rapidly across frameworks, compilers, kernels, network, and hardware
generations~\cite{ncclpaper, fa1,fa2,fa3,fa4, pytorch}. The other
  option is downscaling: running the full training stack on a reduced deployment.
  For some workloads, no viable configuration shrinks the job enough to fit the
  available hardware. Even when one does, reducing the world size and
  reconfiguring parallelism and batch size changes exactly the behaviors
  engineers care about: communication structure, per-GPU workload, memory layout,
  and bottlenecks~\cite{megatron-lm, zero1}. Moreover, scale-dependent effects,
  such as stragglers in synchronous collectives, may disappear. Hence,
conclusions drawn from downscaled experiments can therefore fail to transfer to
production. In short, we lack a way to study large-scale training behavior
faithfully without paying excessive cost.

This paper presents \sysname, a system that enables faithful emulation of
large-scale LLM training using limited hardware resources. The key insight
behind \sysname is that large-scale interaction structure does not require all
ranks---the per-GPU worker processes of a distributed training job---to execute
simultaneously on physical GPUs. Instead, scale can be virtualized by capturing
the execution structure of a training job and selectively replaying it while
running only a subset of ranks on real hardware. By decoupling the logical
training scale from physical GPU allocation, \sysname allows engineers to
observe large-scale behavior while using only a small fraction of the original
hardware.

\sysname achieves this through a two-phase design. First, \sysname constructs a
high-fidelity execution representation \graphname that captures computation,
communication, and their dependencies across all ranks of a training job.
To construct this graph with only a few GPUs, \sysname uses context
  switching to execute many logical ranks in turn. It then runs successive slices
  of ranks to measure operation durations and aligns the resulting timings across
  slices. It can also inject user-specified outlier behavior into the graph.
Second, \sysname performs hybrid emulation in which selected ranks execute the
real training program on physical GPUs while the remaining ranks are replayed
as virtual participants that preserve the communication and synchronization
behavior of the original deployment. This design allows engineers to run
unmodified training code while observing realistic, large-scale interactions.

In summary, this paper makes the following contributions:
\begin{itemize}[leftmargin=*, noitemsep]
  \item We identify faithful emulation of large-scale LLM training as a systems problem
        that neither simulation nor downscaled execution can solve. We show that scale
        can be virtualized: only the ranks of interest need physical GPUs, while the
        remaining ranks participate through replay (\secref{sec:motivation},
        \secref{sec:design}).
  \item We design \sysname's two core subsystems. The \graphsubsysname constructs a
        full-scale \graphname with only a few GPUs through context-switching graph
        collection, slice-by-slice timing filling, inter-slice timing calibration, and
        optional outlier injection (\secref{sec:graph}). The \emulationsubsysname lets
        sandbox ranks run the unmodified program while virtual ranks replay \graphname
        as communication counterparts, made practical by virtual rank initialization
        pruning and collective communication pruning (\secref{sec:emulation}).
  \item We evaluate \sysname on large-scale training workloads of up to 2,048 GPUs
        across diverse models and parallelization strategies, it predicts iteration
        time with 0.58\% average error and peak GPU memory with negligible error.
        Average iteration-time error remains only 0.65\% under straggler
          effects. We further scale up the target scale and show that \sysname can
          emulate clusters of up to 8192 GPUs using fewer than 1\% of the physical GPUs
          required by the original deployment (\secref{sec:evaluation}).
  \item We report deployment experience with \sysname, including training
          configuration tuning on small GPU allocations, memory analysis, targeted
          cluster health checks, extensive profiling, and monitoring tool development
        (\secref{sec:usecases}).
\end{itemize}

\begin{anonsuppress}
  \sysname is available at: \href{https://github.com/AlibabaResearch/CrystalLLM.git}{https://github.com/AlibabaResearch/CrystalLLM.git}.
\end{anonsuppress}

\section{Motivation}
\label{sec:motivation}

\subsection{The Dilemma of DevOps in LLM Training Infrastructure}
\label{subsec:devops}

LLM training infrastructure engineers operate in a DevOps loop. They analyze
telemetry and profiling data from prior or ongoing runs to identify performance
bottlenecks and establish optimization targets (e.g., lower iteration time or
reduced memory overhead). Subsequently, they implement changes spanning the
Python orchestration, distributed runtime, collective communication libraries,
and CUDA kernels. After validation, these changes are rolled out to production
jobs at permitted boundaries such as checkpoints or restarts and rolled back if
regressions occur~\cite{robust_bytedance}.

While engineers can implement and locally validate changes quickly,
small-scale verification provides only partial confidence and
  confirmation often arrives only after large-scale deployment. The performance
  of training infrastructure is determined by intricate interactions across
  mutliple system layers, including the GPU microarchitecture, memory hierarchy,
  and collective communication network. Any of the layers can become the
  bottleneck, and optimizations such as communication--computation
  overlap~\cite{chang2024flux, zhang2025comet} must be designed against this full
  picture. Consequently, small-scale tests often fail to approximate large-scale
  behavior for three reasons. First, the performance of certain layers like the
  collective communication network is inherently sensitive to scale. Second, the
  primary bottleneck can shift across different layers as the system scales
  (e.g., from computation to communication bandwidth or memory
  capacity)~\cite{megatron-lm-large-scale}. Third, some layers harbor latent
  issues that often surface only at scale, such as network contention, high
  memory pressure, and synchronization anomalies~\cite{straggler_bytedance}.

This reliance on large-scale validation faces a fundamental constraint in the
DevOps loop: GPU scarcity. Under the competition of releasing frontier models
every 3--6 months~\cite{epoch_ai_models}, the majority of GPUs are occupied by
production workloads, leaving only a small fraction (hundreds or even tens) for
development, debugging, and validation. Meanwhile, the stack keeps moving:
kernels are frequently redesigned and fused~\cite{fa1,fa2,fa3,fa4}, and new GPU
generations arrive roughly every two years~\cite{epoch_ml_hardware} with
different compute, memory, and interconnect characteristics. Conclusions
therefore age quickly, and each shift sends engineers back to the scarce
large-scale resources for re-validation.

\subsection{Simulating Large-scale Clusters?}
\label{subsec:simulating}

Engineers have long sought methods to evaluate the performance of LLM training
infrastructure without requiring large-scale deployments. Prior work has
explored simulation-based
systems~\cite{astrasim1,astrasim2,simai,phantora,vtrain,multiverse,daydream,dpro,distsim}.
These systems approximate large-scale training behavior by constructing and
replaying intermediate representations using techniques such as execution trace
collection, operator-level profiling, and analytical models, enabling
performance estimation without full-scale execution, and are effective for
early design exploration.

However, simulation accuracy depends on the fidelity of the underlying models:
more detailed models capture more behaviors but take more engineering to build
and maintain, and this trade-off steepens as the stack grows more complex and
evolves faster. SimAI~\cite{simai} maintains a mocked version of the training
framework alongside the real one; Phantora~\cite{phantora}, though simulating
at a lower level, still models each CUDA operator and network behavior by hand.
Some behaviors---GPU kernel scheduling, thermal throttling---are opaque to the
modeler and resist accurate modeling. For questions that require reproducing
production behavior, this modeling cost is paid with every change to the stack.
As the assumptions embedded in a model drift from the evolving system,
profiling and re-implementation recur~\cite{simai, phantora}.

\subsection{Downscaling the Testing Environment?}
\label{subsec:downscaling}

Another option is downscaling: running the full stack on a reduced deployment
to approximate production behavior, typically by shrinking model parameters,
batch sizes, or parallelism degrees (e.g., DP, PP, TP, and
EP)~\cite{megatron-lm, deepspeed} while preserving the original execution paths
as much as possible.

Constructing a faithful downscaled environment, however, requires careful
reconfiguration. Consider observing the pipeline communication between two
middle stages to improve communication--computation overlap: with fewer GPUs
than production, the remaining parallelism dimensions (DP, TP, and EP) must
shrink to preserve the pipeline configuration, and dependent parameters---the
virtual pipeline parallelism (VPP) degree, the global batch size---must be
reconfigured accordingly~\cite{megatron-lm-large-scale}. The result is a
configuration that diverges from deployment: for instance, shrinking the global
batch size below the VPP degree leaves too few microbatches to drive the
pipeline schedule, a case that requires special handling yet never arises
online.

Even when carefully constructed, downscaled experiments may fail to generalize
to large-scale behavior and can even mislead optimization decisions, as system
bottlenecks shift with scale. CPU activation offloading~\cite{zero1, deepspeed}
illustrates this: deciding which activations to offload and when to load them
back trades memory savings against interference with ongoing computation and
communication; because downscaling alters the per-GPU workload, memory layout,
and pipeline schedule, the measured offloading benefit no longer matches
production.

Furthermore, downscaling may mask latent issues that only emerge from
  large-scale interactions. An example is the straggler effect in collective
  communication, a pervasive cause of performance degradation in large-scale
  training~\cite{straggler_bytedance}. Parallelization strategies like DP
  introduce synchronous collective operations at specific points; any variance in
  arrival times across participants forces early-arriving ranks to sit idle
  waiting for the slowest one. Because the existence of stragglers is predicated
  on a large-scale interaction, shrinking the parallelism degree (e.g., reducing
  the DP degree to 1) eliminates the synchronization barrier, making such
  scale-induced anomalies impossible to observe in the downscaling test.

\subsection{Selective GPU Execution: A Promising Idea}
\label{subsec:selective_idea}

Field development workflows suggest a different intuition. When debugging or
analyzing large-scale training workloads, engineers rarely need to inspect the
metrics of all ranks simultaneously. Instead, they typically focus on a small
subset of ranks of interest (e.g., 8 GPUs), while the remaining ranks simply
participate in the distributed execution. For example, engineers often examine
the end-to-end iteration time by observing when the last rank in the pipeline
completes an iteration, or inspect the execution path of a specific rank to
understand program logic or diagnose failures. Similarly, debugging issues such
as out-of-memory (OOM) errors or communication stalls typically requires
detailed inspection of only a few ranks rather than the entire deployment.

This observation raises a potential opportunity: \textit{can we execute only
  the ranks of interest physically while maintaining the rest of the system
  logically?} Such an approach could preserve the deployment-scale interaction
structure while significantly reducing both hardware requirements and
engineering to build and maintain models for simulation. Ideally, we can build
an emulator that decouples logical scale from physical hardware. It
  first records a large-scale workload---capturing what each rank does and
  when---to enable a few GPUs to faithfully reproduce the full-scale behavior.
  Engineers can then specify the ranks of interest to run the real training
  program on physical GPUs, while the remaining ranks are virtualized as logical
  ranks to replay the recorded behavior, thereby preserving the interaction
  structure of the full deployment.

\subsection{Envision: A Trace-driven Emulator}
\label{subsec:goals_challenges}

Based on our in-production experience, we envision the emulation system that
achieves three goals:

\begin{itemize}[leftmargin=*, noitemsep]
  \item \textbf{High fidelity.} The system should faithfully reproduce the behavior of
        the ranks of interest as large-scale runs.
  \item \textbf{Low resource footprint.} The system should emulate the large-scale
        workload with only a few GPUs to reduce hardware requirements.
  \item \textbf{Code reuse.} The emulation system should directly reuse the current
        training code base to avoid unnecessary development or maintenance overhead.
\end{itemize}

A natural approach to building such an emulator is record-and-replay.
If hardware resources were unconstrained, this would be trivial: we
  could simply record the execution trace of every rank at full production scale
  and replay it on an identically sized cluster. However, achieving this under a
  low resource footprint while maintaining high fidelity and code reuse
  introduces two challenges.

\noindentpar{Challenge 1: You need scale to see scale.} Replay presupposes a
faithful representation of large-scale behavior---capturing what runs, in what
order, and for how long. Obtaining such a trace, however, seemingly requires
running the job at full scale, which is precisely the hardware cost we aim to
avoid. If we naively multiplex all ranks onto a few GPUs during recording,
severe resource contention distorts the measured timings, which fail to reflect
the original execution and compromise fidelity.

\noindentpar{Challenge 2: Preserving large-scale behavior at small scale.} Even
given a faithful representation, code reuse dictates that the
  unmodified training program on the ranks of interest will behave realistically
  only if the remaining thousands of logical ranks participate and complete all
  collective communications on time. Hosting thousands of communication
  counterparts on a few GPUs risks contending for and even exhausting their
  limited compute and memory budget, which in turn compromises the fidelity of
  the reproduced behavior on the ranks of interest.

\section{\sysname Design Overview}
\label{sec:design}

Our goal is to build a \textit{\textbf{high-fidelity}} and
\textit{\textbf{low-resource-footprint}} emulation system that allows engineers
to run their \textit{\textbf{existing code}} while physically executing only
the ranks of interest, enabling resource-efficient performance evaluation
alongside the faithful reproduction of large-scale training behavior.

\begin{figure}[tb]
  \centering
  \includegraphics[width=\linewidth]{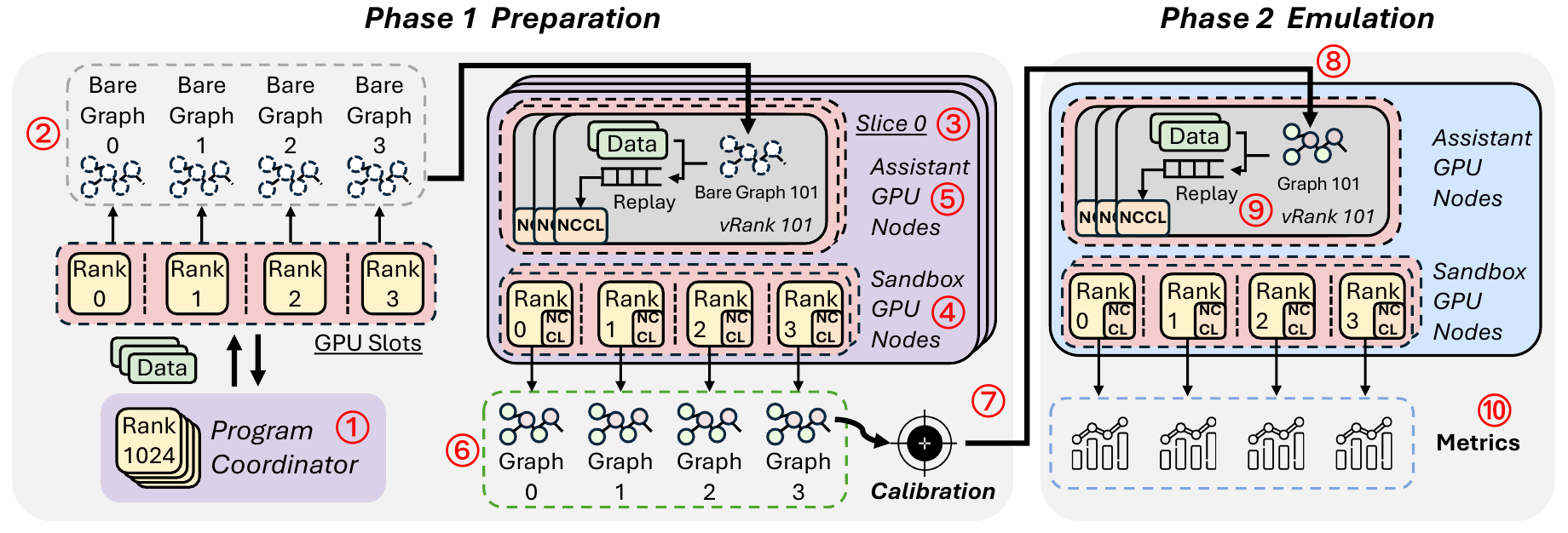}
  \caption{System overview. Preparation collects the bare \graphname, fills in and calibrates its timings. Emulation runs the ranks of interest on sandbox GPUs while virtual ranks (vRank) replay the calibrated \graphname.}
  \Description{Diagram of the two-phase system workflow. Preparation collects a bare execution graph, measures timings in slices, and calibrates them. Emulation runs selected ranks on sandbox GPUs while assistant GPUs host virtual ranks for communication replay.}
  \label{fig:overview}
\end{figure}

We structure \sysname's record-and-replay approach into two phases 
  (Figure~\ref{fig:overview}): \textbf{preparation}, which captures a
  high-fidelity execution representation (\graphname) of the large-scale workload
  using only a few GPUs, and \textbf{emulation}, which executes the unmodified
  training program only on the ranks of interest within the sandbox while
  replaying the captured \graphname around them, as if deployed at full scale.

To overcome the aforementioned challenges
  (\secref{subsec:goals_challenges}) of realizing this workflow at a small scale,
  \sysname comprises two core subsystems: (1) The \textbf{\graphsubsysname
    (\secref{sec:graph})} drives the \textit{preparation} phase by constructing the
  full-scale \graphname on limited hardware through a combination of
  context-switching graph collection, slice-by-slice timing profiling, and
  inter-slice timing calibration. (2) The \textbf{\emulationsubsysname
    (\secref{sec:emulation})} supports both the \textit{preparation} and
  \textit{emulation} phases. It orchestrates lightweight communication
  counterparts to maintain collective communication with the ranks of interest in
  the sandbox, while bounding their resource footprint to prevent contention over
  the limited compute and memory budget.

\subsection{System Workflow}
\label{subsec:workflow}

\textbf{Phase 1: Preparation.} In this phase, the \graphsubsysname constructs a
high-fidelity execution representation (\graphname) for all ranks in the
large-scale training job. \sysname's centralized
\textit{coordinator}~\circled{1} first executes the original program on a small
number of GPUs, scheduling a subset of ranks at a time and coordinating their
progress. It multiplexes execution across ranks and collects a bare
\graphname~\circled{2} that encodes operations and dependencies without
accurate timing.

To obtain accurate timings with limited GPUs, \sysname divides the ranks into
\textit{slices}~\circled{3}, each a subset that can execute natively on the
available sandbox GPUs. It runs the slices sequentially using the
\emulationsubsysname. In each slice, the available GPUs serve two roles:
\textit{sandbox GPUs}~\circled{4} run the ranks in the slice natively, while
\textit{assistant GPUs}~\circled{5} replay virtual ranks as communication
counterparts. The virtual ranks respond to the sandbox ranks' communication
operations (e.g., collectives), allowing the sandbox ranks to observe realistic
communication timing. \sysname iterates over the slices in round-robin order so
that every rank runs natively once.

After collecting per-slice execution timings~\circled{6}, \sysname
  calibrates them across slices to construct a complete \graphname with globally
  consistent timings~\circled{7}. To inject outlier behavior, users can supply
  operation-time distributions; \sysname samples them during calibration to
  assign varied durations across replicas.

\textbf{Phase 2: Emulation.} With the calibrated \graphname, \sysname reuses
the \emulationsubsysname for emulation~\circled{8}. The specified ranks of
interest execute the original program on \textit{sandbox GPUs} and serve as the
primary observation points. \textit{Assistant GPUs} host the virtual ranks
needed to communicate with the sandbox ranks~\circled{9}; these virtual ranks
replay the calibrated \graphname and perform real communication. This phase
runs a single emulation instance for the specified ranks of interest, rather
than iterating over slices. This enables \sysname to measure metrics of
interest (e.g., end-to-end iteration time and GPU memory usage over time) for
sandbox ranks~\circled{10} without requiring full-scale hardware.

The required sandbox size depends on the stage. Collecting the bare graph runs
on any number of GPUs, as timing is discarded at this stage. For
timing-sensitive stages---slice-by-slice timing filling and hybrid
emulation---\sysname provisions the sandbox at node granularity, with enough
nodes to host one tightly-coupled parallelism group
($\max(\mathrm{TP},\mathrm{EP})$ ranks; e.g., two 8-GPU nodes for EP=16). These
groups communicate at per-layer, microsecond cadence (e.g., TP collectives over
NVLink), a granularity at which virtual peers risk perturbing the observed
timing; meanwhile, a sub-node sandbox saves little, as it still needs an
assistant node to carry its inter-node traffic while complicating intra-node
replay. In common configurations, the tightly-coupled group can fit on one
node, and emulation can use as few as two machines: one sandbox node and one
assistant node. The number of assistant GPUs must match that of the sandbox
GPUs to ensure sufficient network bandwidth and memory capacity for the
communication.

\subsection{Assumptions and Scope}
\label{subsec:assumptions}

\noindentpar{Assumptions.} \sysname follows an offline-capture, online-replay
strategy: the execution representation is captured on a small GPU cluster, then
replayed to drive emulation. This strategy holds under three assumptions. (1)
\emph{Deterministic execution.} Given the same program and training
configuration, every run issues the same operation sequence, so graphs captured
for different ranks across runs can be merged into a unified graph for replay.
If the configuration changes, the graph needs to be re-captured. (2) 
  \emph{Replica symmetry.} Because symmetric replicas, such as those in data
  parallelism (DP), generate structurally identical graphs, capturing a single
  model replica suffices; its graph can then be reused across all replicas
  (\secref{subsec:collecting_bare_graph}). Furthermore, the captured timings
  reflect the normal (average) performance of each operation related to the GPUs
  used during timing filling. Since symmetric ranks exhibit statistically
  indistinguishable timing during steady-state pretraining, runtime outliers
  (e.g., stragglers) cannot be natively captured into the graph; such information
  can be injected into the graph as external inputs during calibration instead
  (\secref{subsec:graph_calibration}). (3) \emph{Timely data supply.} Online
replay must keep pace with the ranks of interest in the sandbox: communication
data of virtual ranks must be GPU-resident before ranks in the sandbox reach
the corresponding operation. This holds when the prefetch bandwidth
(host-to-GPU PCIe, e.g., 64~GB/s) exceeds the network bandwidth consuming it
(e.g., 50~GB/s per NIC) (\secref{subsec:emulation_replay}).

\noindentpar{When \sysname is appropriate.} \sysname is appropriate for
training jobs whose execution and communication are deterministic across runs,
supporting mainstream pretraining stacks such as
Megatron-LM~\cite{megatron-lm}, DeepSpeed~\cite{deepspeed}, and
FSDP~\cite{pytorch_fsdp}. It targets questions that can be answered by
observing selected ranks over one or a few iterations. Our prototype currently
supports training code based on PyTorch that communicates through
NCCL~\cite{ncclpaper} (NVIDIA's collective communication library); fused
compute--communication kernels are not yet supported. Extensions to additional
stacks or communication libraries must uphold the same invariant as \sysname
---the exact behavior it would have at full scale must be retained in the
sandbox.

\noindentpar{When not to use \sysname.} \sysname is not the right tool in the
following cases. (1) When runnable code for the feature being evaluated or the
target hardware does not yet exist, simulation is more appropriate because
\sysname relies on actual execution of the target software on target GPUs. (2)
When runtime data values alter the execution path and consequently the executed
operations, as in some RL post-training workloads. Because this violates the
deterministic execution assumption, fixed-graph replay becomes invalid. (3)
When evaluating metrics that require genuine computation by all ranks over many
iterations, including convergence or numerical drift and congestion created
jointly by all traffic. \sysname may identify an actually faulty sandbox node
relative to a normal baseline (\secref{sec:usecases}), but replay does not
synthesize rare failures (e.g., transient stragglers).

\section{Constructing High-Fidelity Representation}
\label{sec:graph}

To enable replay during emulation, the \graphsubsysname of \sysname
  constructs \graphname, a replay-oriented graph representation
  (\secref{subsec:graph_structure}) that captures essential information while
  faithfully preserving large-scale behavior at low cost. It constructs the
  complete \graphname in two stages: first, it collects a bare graph that
  captures execution structure (\secref{subsec:collecting_bare_graph}); then, it
  fills in and calibrates its timing (\secref{subsec:graph_calibration}) to
  ensure accuracy.

\subsection{\graphname: A Replay-Centric Execution Graph}
\label{subsec:graph_structure}

\sysname constructs the \graphname that captures both computation and
communication across all ranks, providing a complete view of large-scale
behavior for replay. For each rank, the graph must answer three questions: (1)
\textit{what} operations are executed (computation and communication spans),
(2) \textit{in what order} they execute (intra- and inter-rank dependencies),
and (3) \textit{how long} they take (duration of each span).

A common approach is to use the PyTorch Profiler~\cite{pytorch} with tools such
as Chakra~\cite{chakra} and Meta HTA~\cite{meta_hta}. However, PyTorch Profiler
incurs significant overhead---we observe up to a 20.04\% increase in iteration
time in a 128-GPU Qwen3 training job~\cite{qwen3}---because it is not designed
for replay. During replay, the graph serves as an input to the
  \emulationsubsysname, which uses it to determine when each virtual rank should
  issue communication instead of re-executing computation. Therefore, we only
  need to preserve the aggregate timing of computation spans between
  communications and their dependencies; fine-grained operator-level details are
  redundant.

\begin{figure}[tb]
  \centering
  \includegraphics[width=\linewidth]{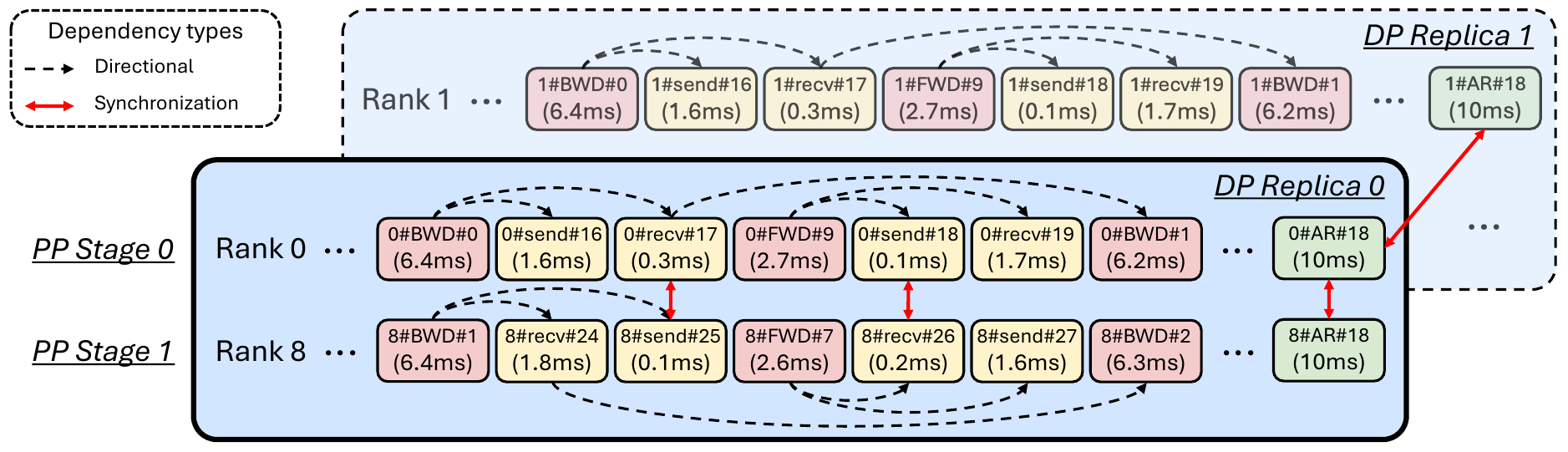}
  \caption{An example snippet of a \graphname capturing execution dependencies of the 1F1B schedule.}
  \Description{A directed dependency graph for a 1F1B pipeline schedule. Each node represents a computation span or a communication event, and directed edges indicate their execution and synchronization dependencies.}
  \label{fig:graph_structure}
\end{figure}

Motivated by this insight, we design \graphname, a new tracing format that
captures only the minimal information required for replay. \graphname operates
at a coarser scheduling granularity (e.g., per microbatch), where each
scheduling unit defines the minimal dependency boundary needed for subsequent
communication.

Figure~\ref{fig:graph_structure} shows an example snippet of a \graphname. It
consists of \textit{nodes} and \textit{edges}. Each node represents a
  computation span or a communication event: it is uniquely identified by an
  \smalltt{id} and records a \smalltt{duration} measured on the GPU with CUDA
  events. Communication nodes additionally store essential metadata (e.g.,
  communication type, group, and participants). Each edge encodes a dependency
between two nodes $(\smalltt{id}_1, \smalltt{id}_2)$. We distinguish two
dependency types: (1) \textit{directional}, where one operation must complete
before another begins (e.g., communication followed by computation within a
rank), and (2) \textit{synchronization}, where all participating nodes must
reach the operation before any can proceed (e.g., collectives or matched
send--receive pairs).

\graphname records timings from the GPU side only, without explicitly tracing
CPU-side events. Because GPU execution determines the actual communication
timing---not the time at which the CPU issues the call---CPU-side timestamps
are unnecessary for replay. CPU-side effects are still captured: bottlenecks
(e.g., data-loading stalls or kernel-launch queuing) manifest as longer gaps
between the GPU events recorded in the graph, and these gaps are therefore
replayed as-is. By focusing exclusively on execution structure and
communication boundaries, \sysname captures only what is necessary for faithful
replay, significantly reducing overhead during graph construction.

\subsection{Stage 1: Collecting a Bare \graphname}
\label{subsec:collecting_bare_graph}

At first glance, constructing the \graphname for all ranks appears to require
running the training job at scale, which conflicts with the requirement of
using only a small number of GPUs. To address this, the \graphsubsysname
introduces context switching to decouple logical execution from physical scale,
enabling bare graph collection on a small number of GPUs.

At a high level, \sysname maintains a pool of logical ranks, each with its own
execution state. At any moment, only $N$ ranks are active on the $N$ available
GPUs, while the remaining ranks are paused in CPU memory (or spilled to disk).
A centralized coordinator schedules batches of $N$ logical ranks onto the GPUs
and runs each rank until it either blocks at a communication boundary or
finishes. When a rank blocks, the coordinator saves its execution state and
communication context, then swaps it out for another runnable rank. By
iteratively advancing this process across all ranks, \sysname collects the bare
graph without requiring large-scale hardware. Reconstruction is needed only
when a change affects the graph's structure, such as changes to parallelization
settings or collective communication metadata; compute- or memory-only changes
(e.g., kernel fusion or recomputation) reuse the bare graph.

\begin{figure}[tb]
  \centering
  \begin{subfigure}[t]{\linewidth}
    \centering
    \includegraphics[width=0.8\linewidth]{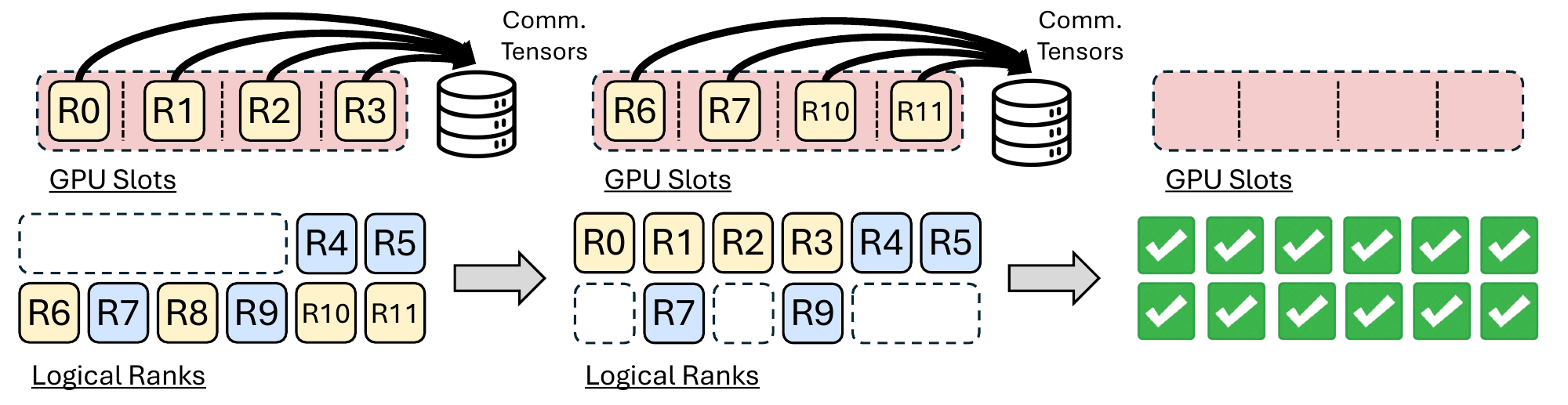}
    \caption{By context switching}
    \label{subfig:context_switching_context_switching}
  \end{subfigure}

  \begin{subfigure}[t]{\linewidth}
    \centering
    \includegraphics[width=0.8\linewidth]{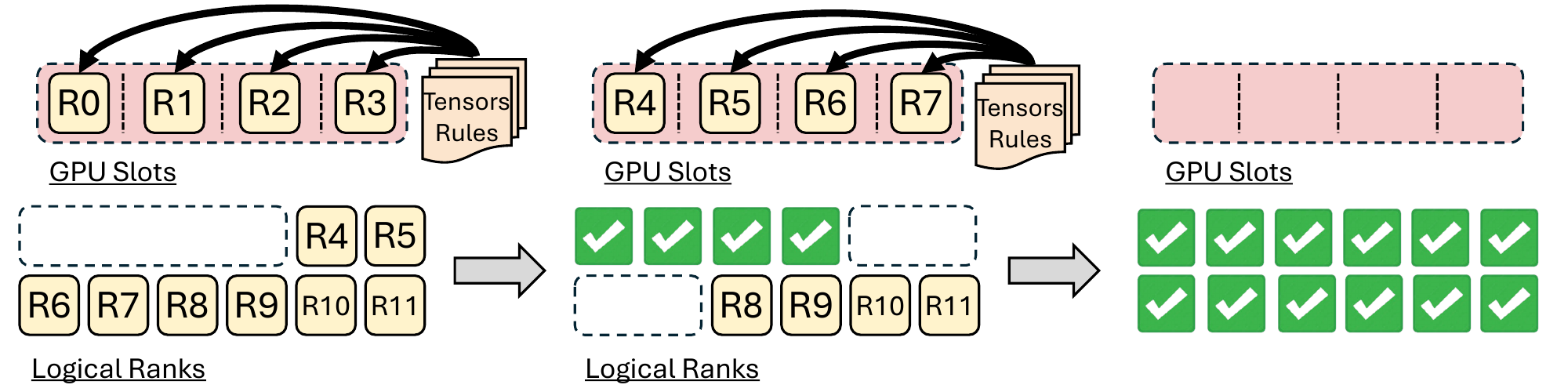}
    \caption{By user-defined communication tensors}
    \label{subfig:context_switching_tensor_rules}
  \end{subfigure}
  \caption{Collecting a bare \graphname with four available GPUs.}
  \Description{Two left-to-right workflows compare bare graph collection for 12 logical ranks using four GPU slots. In the upper workflow, context switching saves communication tensors as rank batches execute; in the lower workflow, supplied tensor rules provide communication inputs so ranks can run independently in batches.}
  \label{fig:context_switching}
\end{figure}

\noindentpar{End-to-end workflow.} \sysname runs unmodified training programs
transparently and manages all ranks through a centralized coordinator. The
input can be any large-scale training job (e.g., one using 256 GPUs with PP=4,
TP=4, and DP=16), regardless of its parallelization configuration.

As shown in Figure~\ref{fig:context_switching}, at the start of each iteration,
the coordinator schedules up to $N$ (e.g., 4) ranks onto the available GPUs and
begins collecting the bare graph for these ranks. Each rank runs until it
reaches a communication point (e.g., an allreduce). The coordinator then checks
whether all participants in the collective are active. If not, the collective
cannot proceed. For example, in
Figure~\ref{subfig:context_switching_context_switching}, ranks R0, R1, R2, R3,
R6, R8, R10, and R11 belong to the same collective, but cannot all be active
simultaneously, so the ranks that reach the collective block. Each blocked rank
stores its communication input tensors in host memory, checkpoints its GPU
context, and swaps out to free the GPU.

Next, the coordinator schedules ranks that can unblock the collective (e.g.,
R6, R8, R10, and R11), prioritizing such ranks to reduce context switching
(Appendix~\secref{app:priority-switching}). Once all participants reach the
collective and their inputs are available, \sysname executes the collective on
CPU and produces outputs. These outputs unblock stalled ranks, which are later
rescheduled and resume execution where they paused.

This process---execute, block, swap out, gather tensors, resolve communication,
and resume---repeats until all ranks complete the iteration. Because each rank
executes the original program with real tensor values, this approach guarantees
that the program follows the correct code paths and handles value-dependent
logic faithfully. By the end, \sysname has collected the full-scale bare
\graphname across all ranks, including the operations (\textit{what}) and their
dependencies (\textit{in what order}). Timing information (\textit{how long})
is added later through timing filling and calibration
(\secref{subsec:graph_calibration}) to construct the complete graph.

\noindentpar{User-defined communication input.} While context switching
correctly handles value-dependent logic, it can incur non-trivial overhead when
GPUs are limited. First, storage overhead can be substantial: an A100 GPU has
80~GB of memory, so replaying a 128-GPU job requires storing up to $128 \times
  80$~GB ($\approx 10$~TB) of data in host memory for swapping. To scale beyond
host memory limits, \sysname can spill inactive contexts to disk using
CRIU~\cite{criu}. Second, GPU context switching itself is expensive:
checkpointing and restoring a near-full 80~GB context can take more than two
minutes with \smalltt{cuda-checkpoint}~\cite{cuda_checkpoint}.

To avoid these overheads, \sysname provides an alternative for advanced users
who have prior knowledge of their workload: they can directly specify the input
and output tensors required for communication, bypassing context switching
entirely. These tensors can be retrieved from communication data recorded
during prior large-scale runs or generated using user-defined rules. Most
tensors, such as weights and gradients, are used for numerical computation
rather than to steer the execution path, and values within a reasonable range
typically suffice. User-defined rules matter for the few tensors whose values
steer the program's execution path. Our Megatron-LM workloads involved three
such types of tensors: \emph{dataloader statuses}---rank~0 broadcasts its
dataloader status after construction, and a status indicating failure would
terminate training, so we inject a ``successful'' status; \emph{training
  samples}---embedding-layer ranks receive token indices via broadcast within
their TP group, and out-of-vocabulary indices trigger index-out-of-bounds
errors, so we inject valid in-vocab values; \emph{MoE dispatch all-to-all
  splits}---an allgather of the gating results determines the size of the
subsequent all-to-all buffers. Overly large splits cause unintended OOMs, so we
inject values that keep the splits within a reasonable range.

With user-provided tensors, each rank can execute independently on the
available GPUs using its corresponding inputs, as shown in
Figure~\ref{subfig:context_switching_tensor_rules}. For a 2,048-GPU Qwen3 235B
MoE training job, collecting the full-scale bare graph on 32 GPUs takes about 9
hours with context switching, whereas using user-defined communication inputs
reduces the collection time to about 64 minutes (64 rounds of roughly one
minute each).

\begin{figure}[tb]
  \centering
  \includegraphics[width=\linewidth]{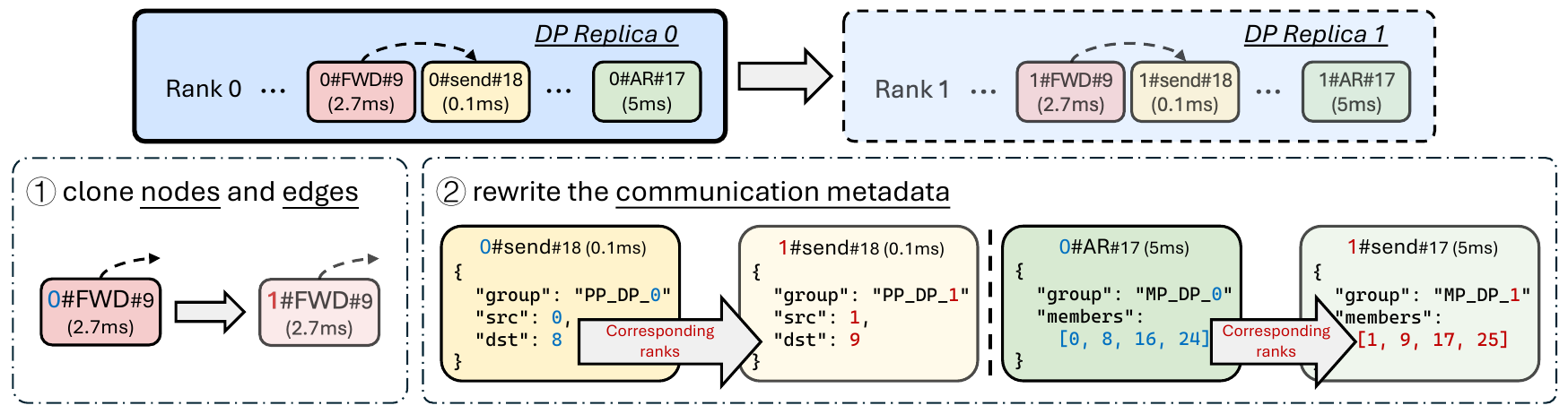}
  \caption{Cloning a bare \graphname and remapping communication metadata across data parallelism replicas.}
  \Description{The upper portion shows matching computation and communication nodes and durations for Rank 0 in DP Replica 0 and Rank 1 in DP Replica 1. The lower portion illustrates two steps. First, nodes and edges are cloned. Second, communication metadata is rewritten.}
  \label{fig:graph_homogeneity}
\end{figure}

\noindentpar{Exploiting parallelism homogeneity.} Decoupling rank
  execution via user-defined inputs unlocks a further optimization: exploiting
  structural \textit{homogeneity} and \textit{heterogeneity} across parallel
  groups to avoid redundant collection. Specifically, parallelization strategies
  such as data parallelism (DP) exhibit homogeneity: corresponding ranks across
  different DP replicas execute identical sequences of operations, yielding
  structurally isomorphic bare graphs that can be reused across replicas. In
  contrast, strategies such as pipeline parallelism (PP) exhibit heterogeneity,
  in which ranks in different stages execute distinct operations (e.g., different
  microbatch schedules) and must therefore be collected individually.
  Consequently, \sysname only needs to collect the structurally heterogeneous
  ranks within a single model replica and can reuse their bare graphs across all
  other homogeneous DP replicas.

However, reusing a bare graph across homogeneous replicas is not a
  direct copy, as communication metadata remains replica-specific. When
  replicating a bare graph for target ranks, \sysname clones the computation
  nodes and preserves the isomorphic dependency edges, while rewriting the
  communication metadata according to the parallelization topology. As shown in
  Figure~\ref{fig:graph_homogeneity}, for each target DP replica, \sysname
  updates the process group identifier and participating ranks for intra-replica
  collectives (e.g., MP allreduce) and remaps the source and destination ranks
  for point-to-point operations (e.g., PP send/receive). This reduces the rounds
  of collection by a factor of $D$, where $D$ is the DP degree. Continuing the
  2,048-GPU Qwen3 example ($D=64$), \sysname needs to collect only one 32-rank
  model replica in a single round, reducing bare graph collection time from about
  64 minutes to roughly one minute.

\subsection{Stage 2: Timing Filling and Calibration}
\label{subsec:graph_calibration}

\begin{figure}[tb]
  \centering
  \includegraphics[width=\linewidth]{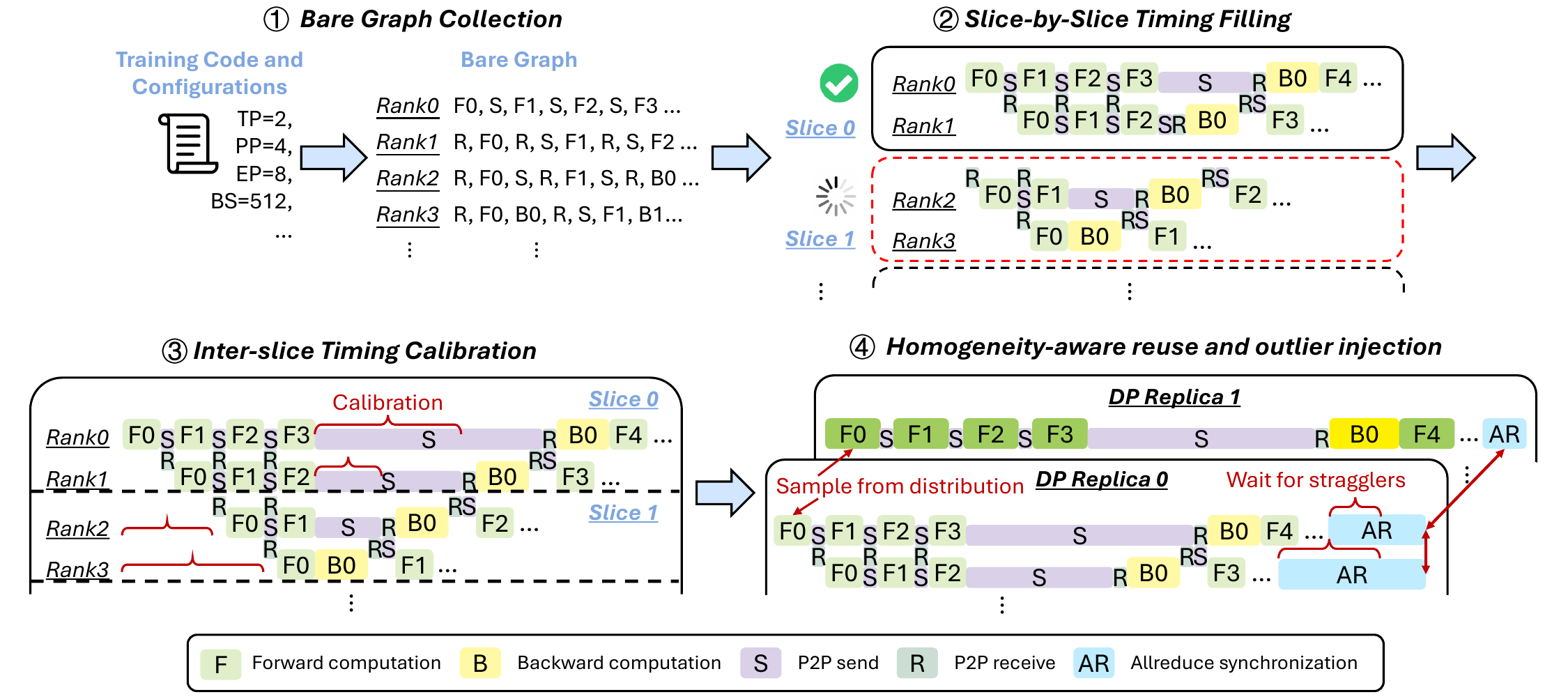}
  \caption{Workflow for timing filling, calibration, and outlier injection.}
  \Description{A four-panel diagram shows how \sysname constructs a complete graph. (1) Training code and parallelization settings produce per-rank bare operation sequences. (2) Ranks are divided into slices; ranks of interest in the active slice execute in the sandbox, while the remaining ranks are replayed virtually using the bare graph. (3) Cross-slice communication dependencies guide calibration to align the independently measured timelines. (4) Timings are reused across homogeneous data parallelism replicas, while outlier durations are sampled and injected; other ranks then wait at an all-reduce for the straggler.}
  \label{fig:calibration}
\end{figure}

After collecting the bare \graphname, we know \textit{what} operations execute
and \textit{in what order}, but not \textit{how long} they take.
Because bare graph collection introduces substantial overhead (e.g.,
  context switching) that distorts the timings of both computation and
  communication, we must measure operation durations in an interference-free
  environment.

To recover accurate timings with limited GPUs, the \graphsubsysname
  leverages the \emulationsubsysname (\secref{sec:emulation}) and divides the
  ranks of the full-scale training job into subsets, or \textit{slices}. As shown
  in Figure~\ref{fig:calibration}, we execute these slices sequentially: for each
  slice, its constituent ranks are designated as the ranks of interest and run
  inside the sandbox with real computation and communication. After all slices
are executed, every rank has run natively in the sandbox at least once,
yielding accurate slice-level timings. We then merge these
  slice-level timings and calibrate them---guided by the parallelization and
  scheduling semantics encoded in the graph's dependencies---to construct a
  complete, globally consistent \graphname. Furthermore, we again exploit
  structural homogeneity and heterogeneity to accelerate this timing filling
  process, and build upon it to inject runtime outlier information into the
  graph.

\noindentpar{Slice-by-slice timing filling.} In each slice execution, the
designated ranks of interest run the unmodified program natively in the
sandbox, while the remaining ranks are replayed as virtual ranks on assistant
GPUs using the bare graph. For a 1,024-rank job, for instance, one slice
executes ranks 0--7 in the sandbox while replaying ranks 8--1,023 on assistant
GPUs; \sysname steps through all slices in a round-robin manner (e.g., 0--7,
8--15, \dots), so that every rank runs natively once.

Because the ranks in the sandbox execute natively, intricate runtime dynamics
such as multi-stream computation--communication overlapping are faithfully
preserved. Virtual ranks serve as communication counterparts through the same
collective stack, feeding the communication tensors recorded or specified in
Stage~1. Consequently, each rank observes realistic network interactions,
yielding accurate computation and communication timings.

\noindentpar{Inter-slice timing calibration.} Because slices are run
independently, the measured timings are locally accurate within each slice but
lack global alignment across slices. \sysname therefore leverages the
inter-rank communication dependencies encoded in the graph to calibrate timings
across slice boundaries.

For example, Figure~\ref{fig:calibration}~\circled{3} illustrates a 4-stage
pipeline-parallel job partitioned into two slices (slice~0 and slice~1), each
yielding locally accurate node durations. To correlate them, \sysname leverages
cross-slice dependency constraints: because a \smalltt{recv} operation in
slice~1 depends on its corresponding \smalltt{send} in slice~0, \sysname aligns
their execution timelines by shifting the \smalltt{recv} to complete only after
the \smalltt{send} initiates. Propagating these directional and synchronization
constraints across the entire graph reconstructs globally consistent cross-rank
timings and eliminates artifacts introduced by sliced execution. After
calibration, \sysname produces a complete, high-fidelity \graphname that
reflects full-scale behavior.

\noindentpar{Homogeneity-aware reuse and outlier injection.} Similar
  to Stage~1, \sysname accelerates round-robin slice execution by exploiting
  parallelism homogeneity. In normal large-scale deployments, symmetric ranks
  across homogeneous replicas (e.g., DP groups) exhibit statistically
  indistinguishable timings during steady-state pretraining. Consequently, rather
  than running all slices across the entire job, \sysname performs timing filling
  only for slices within a single model replica and reuses these measured timings
  across all other homogeneous replicas.

However, the timings measured during timing filling reflect the
  average performance of the specific physical GPUs used for the sandbox. In
  real-world large-scale deployments, runtime outliers---such as stragglers
  caused by hardware degradation~\cite{straggler_bytedance}---cannot be natively
  captured, as the small set of sandbox GPUs rarely encompasses both healthy and
  degraded hardware simultaneously. Yet, when coupled with synchronous
  collectives (e.g., DP all-reduces), even a single slow rank can bottleneck the
  entire training job. To emulate such scenarios, \sysname allows injecting
  outlier behavior as an external input during calibration as shown in
  Figure~\ref{fig:calibration}~\circled{4}. Specifically, users can provide a
  distribution of execution times for target operations, either derived from
  production traces or designed synthetically for what-if analysis. During
  calibration, \sysname samples from this distribution to assign varied durations
  to corresponding operations across different replicas. These perturbed
  durations propagate along the directional and synchronization dependency edges,
  causing delays to cascade into downstream synchronous operations and the
  resulting \graphname captures full-scale execution under straggler effects.

\section{Hybrid Emulation Runtime}
\label{sec:emulation}

The \emulationsubsysname of \sysname supports both timing filling
(\secref{sec:graph}) and final performance evaluation on limited hardware. With
the complete \graphname constructed, \sysname performs hybrid emulation to
measure target performance metrics. We first describe graph-driven replay and
tensor management (\secref{subsec:emulation_replay}), followed by two designs
that bound assistant-GPU overhead: virtual rank initialization pruning
(\secref{subsec:virtualize_init}) and collective communication pruning
(\secref{subsec:virtualize_comm}).

\subsection{Graph-Driven Communication Replay}
\label{subsec:emulation_replay}

\begin{figure}[tb]
  \centering
  \includegraphics[width=0.8\linewidth]{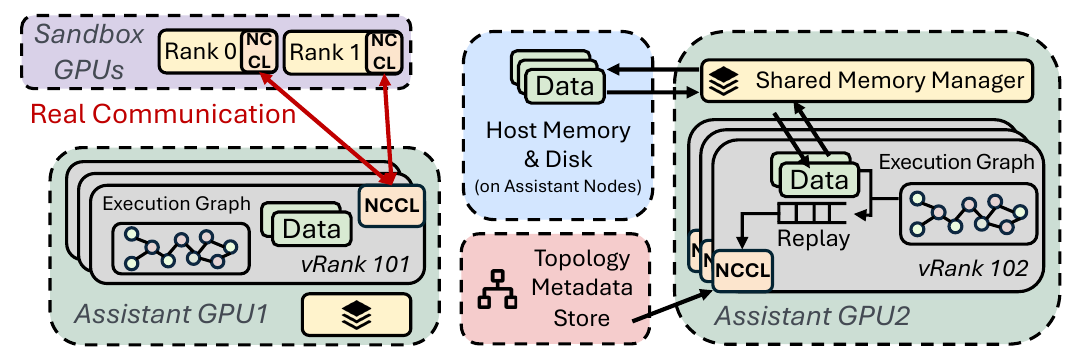}
  \caption{\emulationsubsysname architecture. Sandbox ranks run natively; virtual ranks on assistant GPUs replay the \graphname, communicate with the sandbox, and share tensors through a memory manager.}
  \Description{Architecture diagram showing sandbox ranks communicating with virtual ranks on assistant GPUs. Each virtual rank replays an execution graph and receives communication tensors from a shared memory manager. Host memory and disk on the assistant nodes provide backing storage for these tensors. A topology metadata store supplies NCCL setup information.}
  \label{fig:emulation}
\end{figure}

\noindentpar{Graph-driven replay.} Figure~\ref{fig:emulation} illustrates the
runtime. During emulation, the designated ranks of interest execute the
unmodified training program natively on \textit{sandbox GPUs}, while
\textit{assistant GPUs} host virtual ranks that provide their communication
counterparts, multiplexing multiple virtual ranks onto each physical assistant
GPU. To conserve resources, initialization instantiates only virtual ranks that
communicate directly with the sandbox GPUs (see
\secref{subsec:virtualize_init}). During replay, each virtual rank traverses
its pre-constructed \graphname. At a computation span (e.g.,
\smalltt{\#0\#FWD\#1}), it bypasses real computation and waits for the recorded
duration; at a communication node (e.g., \smalltt{\#0\#send\#2}), it issues the
corresponding operation to interact with the sandbox ranks.

\noindentpar{Tensor sharing among virtual ranks.} As multiple virtual
  ranks share a physical assistant GPU, allocating communication tensors
  independently for each virtual rank would duplicate memory and risk exhausting
  GPU memory. Since virtual ranks bypass computation, most tensors only need
  valid shapes, data types, and reasonable values at the time of communication.
  As described in \secref{subsec:collecting_bare_graph}, most tensors (e.g.,
  weights and gradients) do not affect the sandbox ranks' execution paths; only a
  few control-flow-sensitive tensors require exact values. This distinction
  motivates a \textit{Shared Memory Manager} on each assistant GPU. The manager
  maps the virtual ranks' communication tensor requirements to physical storage,
  aggregates requirements across execution graphs of all co-located virtual
  ranks, and orchestrates tensor allocation, reuse, and residency.

For value-insensitive tensors, the manager allocates one shared GPU
  memory pool per data type and initializes it with valid values. Through CUDA
  inter-process communication (IPC), it exposes these allocations to local
  virtual rank processes, which slice tensor views of the required shapes from
  the pools. A uniform \textit{Tensor Provider Interface} handles value-sensitive
  communication tensor requirements: it matches communication operations and
  populates their tensors on demand using captured tensors or user-defined
  rules.

\noindentpar{Tensor residency and prefetching.} Based on the
  communication operations recorded in \graphname, the manager prepares reusable
  buffers and populates the corresponding value-sensitive communication tensors
  before replay. This keeps buffer allocation and data loading off the critical
  communication path. GPU memory holds the active working set, while host (CPU)
  memory serves as a staging and backing tier. If all required tensors cannot
  remain in GPU memory, the manager evicts currently unused tensors to host
  memory and asynchronously prefetches tensors needed for upcoming communications
  during communication-idle intervals, avoiding bandwidth contention.

As the manager maintains tensors only for communications that directly involve
sandbox ranks, the overall buffering footprint is significantly reduced. In
practice, assistant GPU memory is typically sufficient for the entire working
set. For example, when emulating a 512-GPU job with an 8-GPU assistant node,
over 70\% of each GPU's memory remains available for tensor buffering after
accounting for CUDA and NCCL contexts. With lower-memory GPUs, the same
aggregate buffer capacity can be provided by adding assistant GPUs instead.

Under the timely data supply assumption (\secref{subsec:assumptions})---that
PCIe prefetch bandwidth exceeds the rate at which communication consumes the
data---the manager can keep each tensor GPU-resident before its communication
begins. This allows a physical assistant GPU to serve multiple virtual ranks on
demand.

\subsection{Virtual Rank Initialization Pruning}
\label{subsec:virtualize_init}

\begin{figure}[tb]
  \centering
  \begin{subfigure}[t]{0.49\linewidth}
    \centering
    \includegraphics[width=0.5\linewidth]{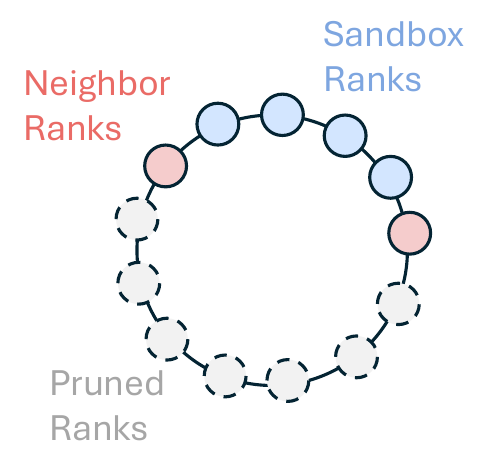}
    \caption{Ring algorithm.}
    \label{subfig:init_pruning_ring}
  \end{subfigure}
  \hfill
  \begin{subfigure}[t]{0.49\linewidth}
    \centering
    \includegraphics[width=0.5\linewidth]{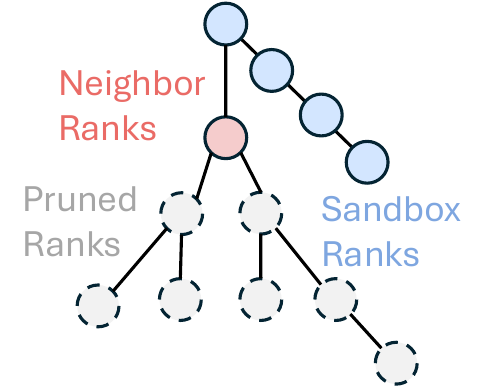}
    \caption{Tree algorithm.}
    \label{subfig:init_pruning_tree}
  \end{subfigure}
  \caption{Virtual rank initialization pruning: only virtual ranks adjacent to the sandbox will be initialized.}
  \Description{Two panels compare ring and tree collective algorithms. In both, only virtual ranks adjacent to the sandbox will be initialized.}
  \label{fig:init_pruning}
\end{figure}

Hosting thousands of virtual ranks as separate processes on a few assistant
GPUs would exhaust their limited compute and memory resources. This overhead
has two main sources. First, large-scale parallel training creates many process
groups (e.g., DP, TP, PP, and EP groups), each requiring communicator state,
transport connections, and buffers across its member ranks. Second, placing
these processes on a few GPUs intensifies contention for CUDA contexts and
driver resources, making initialization partly serialized and potentially hours
long.

\sysname addresses this challenge through virtual rank initialization
  pruning at both process groups and group members levels. This pruning spans
  PyTorch's distributed module and NCCL's initialization path. During
  initialization, sandbox ranks initialize normally, while assistant GPUs
  instantiate only the virtual ranks and communication resources needed to
  communicate with them.

\noindentpar{Process-group pruning.} For virtual ranks, \sysname intercepts
process group creation at PyTorch's \smalltt{c10d} layer and instantiates only
groups whose membership overlaps with sandbox ranks. It bypasses any process
group that does not involve a sandbox rank. In a job with 2,000 virtual ranks,
this filtering reduces the number of active process groups from 8,617 to 82.

\noindentpar{Group-member pruning.} Filtering process groups still leaves
unnecessary members in the retained groups. As shown in
Figure~\ref{fig:init_pruning}, sandbox ranks communicate directly only with
their algorithmic neighbors (e.g., adjacent ranks in a ring, or parent and
child ranks in a tree). Thus, \sysname can prune non-neighboring virtual ranks,
as long as numerical correctness is preserved (see
\secref{subsec:virtualize_comm}), reducing initialization cost and avoiding
unnecessary consumption of communication bandwidth at runtime.

However, Pruning non-neighboring ranks requires overcoming standard
  initialization barriers, which assume that every logical rank is launched. NCCL
  determines a rank's topological neighbors (e.g., ring predecessors and
  successors, or tree parents and children) only after the processes launch and
  exchange hardware information. To select virtual ranks in advance, \sysname
  runs a lightweight offline preparation pass. Assuming cluster homogeneity, this
  pass profiles the hardware topology of one physical node and replicates it
  across the participating logical nodes. It then executes NCCL's graph-search
  routines in a local multithreaded process to construct each group's full
  logical topology before launching actual processes. From this topology,
  \sysname determines which virtual ranks to launch: the sandbox ranks' immediate
  collective neighbors and point-to-point peers, plus essential initialization
  coordinators such as the root rank. It caches this participant plan and the
  precomputed per-rank topology metadata in a centralized store for use during
  initialization. This metadata is associated with the corresponding process
  group identifier, enabling this pruning for \emph{all} process groups that
  perform collective communication.

During initialization, processes entering PyTorch's
\smalltt{init\_process\_group} normally wait at a \smalltt{TCPStore}-based
barrier until every rank has joined. Since pruned ranks are never launched, the
barrier would otherwise wait indefinitely. \sysname avoids this
  without changing user-level world sizes by initializing the underlying
  \smalltt{TCPStore} with the number of launched participants (the sandbox ranks
  plus the selected virtual ranks), while passing each process its original
  logical rank and the original world size. This separates the synchronization
  barrier from the logical group size, allowing framework rendezvous to complete
  once all active processes have joined. Furthermore, NCCL's bootstrap protocol
  still requires hardware and address records for every logical member to
  assemble the communicator. To provide metadata for pruned ranks without
  launching their processes, a designated virtual rank spawns lightweight
  control-plane proxy threads on behalf of all pruned ranks. These proxies
  retrieve the corresponding cached metadata from the metadata store using the
  process group identifier and supply the missing ranks' topology data through
  NCCL's bootstrap control channel. This lets NCCL establish the full logical
  communicator while creating transport connections only for active neighbors.

\subsection{Collective Communication Pruning}
\label{subsec:virtualize_comm}

\begin{figure}[tb]
  \centering
  \includegraphics[width=0.9\linewidth]{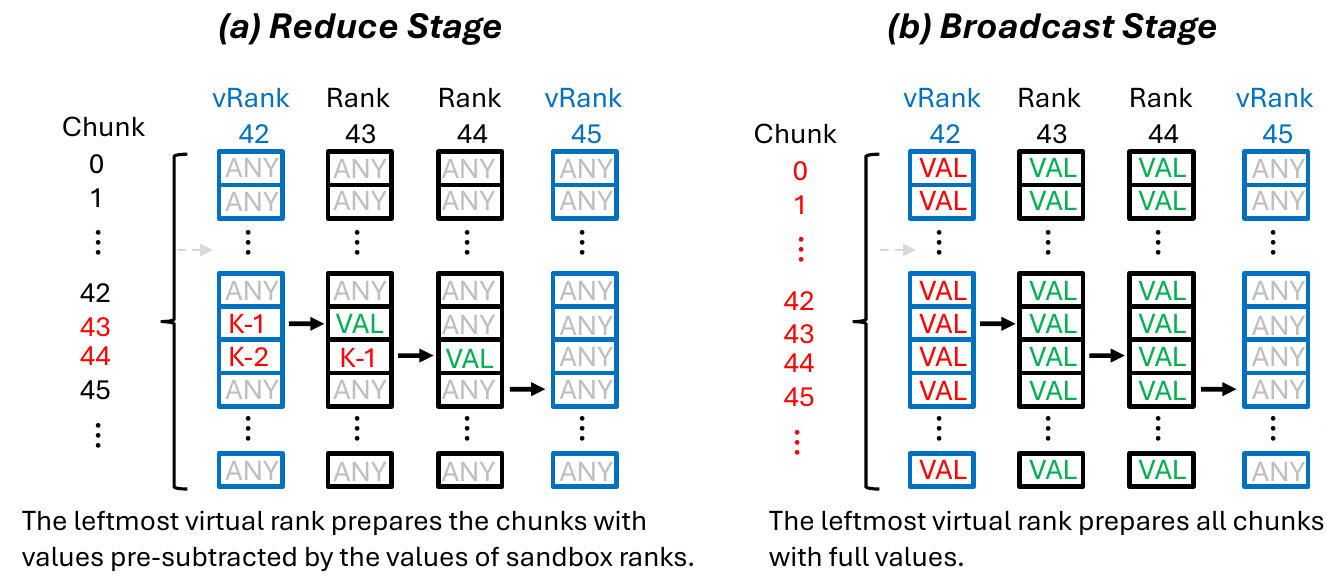}
  \caption{Collective numerical correctness under pruning. Each row is a chunk. (a) Reduce Stage: the leftmost virtual rank (vRank) pre-subtracts sandbox contributions ($K-1$, $K-2$) so sandbox-owned chunks accumulate correct full values (VAL); other chunks carry arbitrary values (ANY). (b) Broadcast Stage: the leftmost virtual rank (vRank) supplies full values for all chunks.}
  \Description{Two panels show reduce and broadcast stages of a ring allreduce after pruning non-neighbor virtual ranks. The reduce stage accumulates correct values for sandbox-owned chunks; the broadcast stage supplies full values for every chunk.}
  \label{fig:pruning_correctness}
\end{figure}

Since we remove non-neighboring ranks, the original collective communication
pattern is changed at runtime. We must therefore ensure that the semantics of
collective operations remain correct under this transformation; to this end, we
modify NCCL's data transmission logic to guarantee numerical correctness.

Most NCCL collectives build on two primitive patterns: \textit{reduction},
which aggregates values toward a destination, and \textit{broadcast}, which
propagates values outward~\cite{ncclpaper}; the remaining operations, such as
all-to-all and point-to-point send/receive, move data directly without
arithmetic. Ensuring correctness for reduction and broadcast therefore covers
the collectives in our workloads.

We illustrate our approach using ring allreduce~\cite{ncclpaper}, which
consists of both \textit{reduction} and \textit{broadcast} stages; the same
principles apply to other collectives. In the reduction stage, each rank splits
its data into multiple chunks and is responsible for aggregating one chunk
(e.g., sum, max, or min) from all other ranks. Let $\mathrm{data}_j^{(i)}$
denote the contribution of rank $j$ to chunk $i$, and let
$\mathrm{data}_{\mathrm{full}}^{(i)} = \sum_j \mathrm{data}_j^{(i)}$ denote the
fully reduced value. The key property is that each rank only needs to ensure
the correctness of its assigned chunk. After one full ring round of $(K-1)$
communication steps, where $K$ is the number of ranks in the ring, each rank
holds $\mathrm{data}_{\mathrm{full}}^{(i)}$ for its assigned chunk $i$. In the
broadcast stage, each rank then propagates its completed chunk to all other
ranks through another ring round $(K-1)$, ensuring that all ranks obtain the
full result.

Figure~\ref{fig:pruning_correctness} illustrates how \sysname preserves
collective correctness under rank pruning. The example includes neighboring
virtual ranks (e.g., \textit{vRank} 42 and \textit{vRank} 45) and sandbox ranks
(e.g., \textit{Rank} 43 and \textit{Rank} 44); other pruned non-neighboring
ranks are omitted for clarity. In the reduction stage, we leverage the key
property of ring allreduce: each rank is responsible only for its assigned
chunk. Therefore, after pruning, we only need to ensure that sandbox ranks
compute their assigned chunks correctly. Once the ranks of interest (e.g.,
\textit{Rank} 43 and \textit{Rank} 44) obtain the correct values for their
chunks, i.e., $\mathrm{data}_{\mathrm{full}}^{(43)}$ and
$\mathrm{data}_{\mathrm{full}}^{(44)}$, the state after the reduction stage is
equivalent to that of the original execution, and the values of other chunks do
not affect the outcome.

To achieve this, the leftmost virtual rank prepares adjusted values that
compensate for missing contributions from pruned ranks. For each chunk, the
goal is to ensure that the value received at the owning sandbox rank matches
$\mathrm{data}_{\mathrm{full}}^{(i)}$. For the chunk owned by rank 44, since
the ring path reaches rank 44 via rank 43, the leftmost virtual rank prepares
$\mathrm{data}_{\mathrm{full}}^{(44)} - \mathrm{data}_{43}^{(44)} -
  \mathrm{data}_{44}^{(44)}$, so that as the data propagates, rank 43 adds back
$\mathrm{data}_{43}^{(44)}$ and rank 44 subsequently adds back
$\mathrm{data}_{44}^{(44)}$, reconstructing the correct final value
$\mathrm{data}_{\mathrm{full}}^{(44)}$. Similarly, for other sandbox-owned
chunks, we subtract the contributions of ranks along the remaining path before
reaching the owner (including the owner itself). For chunks not owned by
sandbox ranks, their values do not affect the final outcome observed in the
sandbox and can therefore be assigned arbitrary values.

In the broadcast stage, each rank propagates its completed chunk to all other
ranks. At this point, the leftmost virtual rank needs to provide correct final
values for all other chunks not owned by the sandbox ranks. Since sandbox ranks
(e.g., \textit{Rank} 43 and \textit{Rank} 44) receive the fully reduced values
$\mathrm{data}_{\mathrm{full}}^{(i)}$ for their respective chunks during this
stage, ensuring correctness for these chunks is sufficient to guarantee overall
correctness from the sandbox’s perspective.

Tree-based algorithms~\cite{ncclpaper} follow the same principle. NCCL builds
hierarchical trees for inter-node communication, with intra-node ranks forming
a chain behind a head rank that carries the node's inter-node traffic;
correctness is again enforced at the boundary where a sandbox head rank meets a
virtual rank. For a given data chunk, the sandbox head plays one of three
roles: 1)~\textit{Root}: the chunk's final reduction completes here, so its
child virtual rank sends a compensated value that, combined with the sandbox's
own contribution, yields the correct result; the subsequent broadcast then
propagates it downward. 2)~\textit{Leaf}: the data it sends upward is reduced
elsewhere and does not affect the sandbox, so correctness relies on its parent
virtual rank returning the correct final value during broadcast.
3)~\textit{Intermediate}: its local reduction is later overwritten by the
broadcast, so its child virtual rank can send arbitrary values while its parent
supplies the correct final value. As with the ring, these boundary rules
preserve numerical correctness while making broad use of arbitrary values.

Beyond numerical correctness, pruning introduces no non-determinism: sandbox
NCCL executes the same number of transmissions with the same volume as the
original collective, and the placement of sandbox and virtual ranks is static
across runs. Repeated emulations therefore differ only by inherent hardware
jitter, comparable to the run-to-run variance of real executions.

\section{Implementation}
\label{sec:implementation}

\sysname uses a \textit{centralized coordinator} to collect a bare \graphname
and \textit{virtual ranks replayers} to replay the graph during slice-by-slice
timing filling and, after calibration, final emulation.

The coordinator orchestrates all logical ranks during bare graph collection. It
runs unmodified training programs and schedules logical ranks on a limited set
of GPUs through context switching. We patch PyTorch's \smalltt{c10d} layer to
intercept communication and execute collectives on CPU without requiring all
participants to be active simultaneously. A CPU collective executor built on
\smalltt{torch.distributed} processes the inputs once all required tensors are
available, then stores the outputs for use when the ranks resume. This
preserves correctness without running all ranks concurrently.

The \emulationsubsysname supports both slice-by-slice timing filling and final
emulation. During timing filling, virtual ranks replay the structure of the
bare \graphname without calibrated timings, while the ranks of interest run
natively on sandbox GPUs; inter-slice calibration aligns the timings measured
across slices. In contrast during final emulation, virtual ranks replay the
calibrated \graphname, including its operation durations.

We also modify NCCL to support collective communication pruning. During
initialization, virtual ranks embed a magic number in their
\smalltt{host\_hash} so that active ranks can identify virtual peers when
exchanging connection metadata. During execution, NCCL skips data transfers on
channels outside the communication paths to sandbox ranks and instead generates
completion metadata to satisfy the collective, reducing overhead while
preserving correctness for the sandbox ranks.

\section{Evaluation}
\label{sec:evaluation}

We evaluate \sysname across four dimensions. First, end-to-end results show
that \sysname achieves high-fidelity predictions for both iteration time and
memory allocation, with average errors of 0.58\% and below 0.01\%, respectively
(\secref{subsec:evaluation_e2e}). Average iteration-time error
  remains only 0.65\% under straggler effects. Second, we evaluate emulation
  efficiency and scalability across target configurations
  (\secref{subsec:evaluation_efficiency}),showing that \sysname can emulate
  clusters of up to 8192 GPUs using fewer than 1\% of the physical GPUs required
  by the original deployment. Third, we conduct microbenchmarks to assess
execution fidelity and quantify the benefits of our key
optimizations—initialization acceleration via virtual rank initialization
pruning, transmission reducction via colletive communication pruning, and
timing alignment via inter-slice timing calibration
(\secref{subsec:evaluation_micro}). Finally, we compare \sysname with
state-of-the-art simulators, including Phantora~\cite{phantora} and
SimAI~\cite{simai}, showing that \sysname achieves higher estimation accuracy
for complex training workloads (\secref{subsec:evaluation_simulators}).

\noindent\textbf{Experimental setup.} We conduct experiments on a 2,048-GPU
testbed (8 GPUs per node) using the open-source Megatron-LM implementation of
Qwen3 MoE pretraining~\cite{pai_megatron_patch}. We evaluate three model
configurations (M.1--M.3) spanning 235B to 1.01T parameters
(Table~\ref{tab:model_structures}).

To cover diverse parallelization strategies under memory constraints, we vary
tensor parallelism (TP), pipeline parallelism (PP), and expert parallelism (EP)
sizes and gradient accumulation steps, yielding four configurations (S.A--S.D,
Table~\ref{tab:parallel_strategies}).

\begin{table}[tb]
  \centering
  \footnotesize
  \begin{minipage}[t]{0.54\linewidth}
    \centering
    \captionsetup{type=table}
    \caption{Model structures evaluated.}
    \label{tab:model_structures}
    \begin{tabular}{@{} l l c c c @{}}
      \toprule
      \textbf{Model} & \textbf{Params} & \textbf{Layers} & \textbf{Heads} & \textbf{Experts (TopK)} \\
      \midrule
      M.1            & 235B-A22B       & 94              & 64             & 128 (8)                 \\
      M.2            & 503B-A20B       & 62              & 32             & 256 (8)                 \\
      M.3            & 1.01T-A43B      & 62              & 64             & 256 (8)                 \\
      \bottomrule
    \end{tabular}
  \end{minipage}
  \hfill
  \begin{minipage}[t]{0.43\linewidth}
    \centering
    \captionsetup{type=table}
    \caption{Parallelization strategies evaluated.}
    \label{tab:parallel_strategies}
    \begin{tabular}{@{}lccccc@{}}
      \toprule
      \textbf{Strategy} & \textbf{TP} & \textbf{PP} & \textbf{VPP} & \textbf{EP} & \textbf{GA} \\
      \midrule
      A                 & 1           & 4           & 0            & 8           & 8           \\
      B                 & 2           & 4           & 2            & 8           & 16          \\
      C                 & 1           & 16          & 0            & 8           & 32          \\
      D                 & 1           & 8           & 0            & 16          & 16          \\
      \bottomrule
    \end{tabular}
  \end{minipage} \par
  \begin{minipage}{\linewidth}
    \vspace{0.5em}
    \textit{*TP: Tensor Parallelism,
      PP: Pipeline Parallelism,
      VPP: Virtual Pipeline Parallelism (number of virtual stages per pipeline stage; 0 means disabled),
      EP: Expert Parallelism,
      GA: Gradient Accumulation.
      Megatron Core distributed optimizer enabled in all setups}
  \end{minipage}
\end{table}

\subsection{End-to-end Metrics Prediction Accuracy}
\label{subsec:evaluation_e2e}

\begin{figure}[tb]
  \centering
  \includegraphics[width=\linewidth]{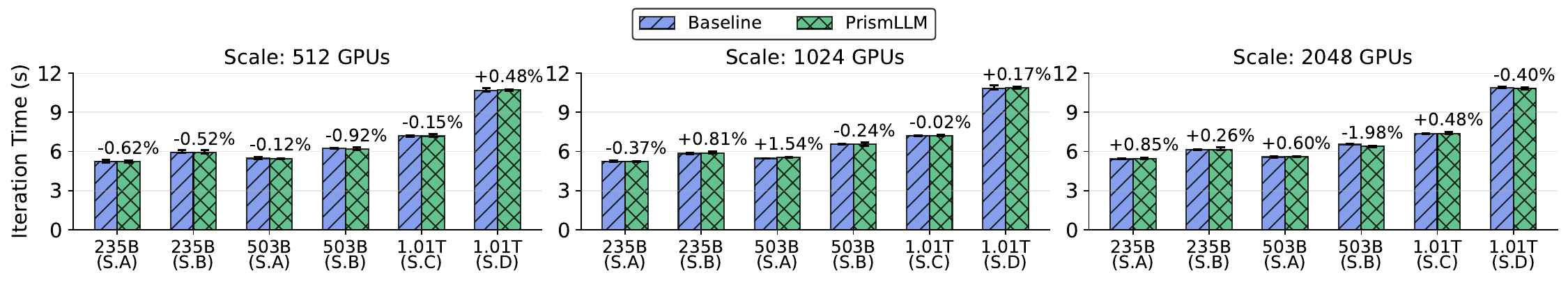}
  \caption{End-to-end iteration time estimation results.}
  \Description{Chart comparing measured baseline iteration times with system estimates across cluster scales, model configurations, and parallelization strategies. The estimates closely track the baseline.}
  \label{fig:experiment_e2e_iter_time}
\end{figure}

\noindentpar{End-to-end iteration time estimation.} We evaluate three cluster
scales (512, 1,024, and 2,048 GPUs), each with three model configurations and
two parallelization strategies. Iteration time is the framework-reported
duration of one step, including computation and communication.

Each test runs the pretraining code at the full target scale for 85 iterations,
and we report the median of the middle 70, excluding warm-up. \sysname emulates
the same workload with identical configurations and scripts. In the
experiments, we used 1 sandbox node and 1, 2, and 3 assistant nodes for
235B(M.1), 503B(M.2), and 1.01T(M.3) with S.C at 512, 1,024, and 2,048 target
GPUs, respectively. For 1.01T(M.3) with S.D, we used 2 sandbox nodes with 2
assistant nodes at 512 GPUs and 4 at larger scales. Simulators have inherent
limits on these workloads (e.g., missing memory-allocation simulation), so we
compare against them separately in \secref{subsec:evaluation_simulators};
downscaled runs are no substitute either, as shrinking the DP degree changes
distributed-optimizer sharding and triggers out-of-memory (OOM) errors.

As shown in Figure~\ref{fig:experiment_e2e_iter_time}, \sysname tracks the
baseline across all scales, with an average error of 0.58\% and a maximum of
1.98\% under diverse parallelization strategies. The use of actual GPUs for
task execution, coupled with the high-fidelity \graphname, empowers \sysname to
achieve highly accurate latency estimations.

\begin{figure}[tb]
  \centering
  \includegraphics[width=\linewidth]{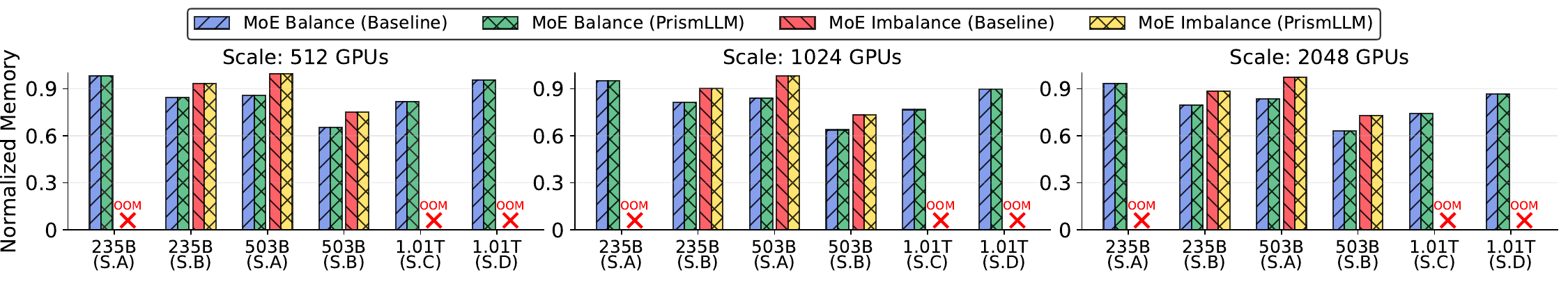}
  \caption{End-to-end peak memory allocation estimation results.}
  \Description{Chart comparing measured and estimated peak GPU memory for balanced and imbalanced workloads at multiple scales. The estimates closely track the baseline.}
  \label{fig:experiment_e2e_memory}
\end{figure}

\noindentpar{End-to-end peak memory allocation estimation.} We evaluate peak
memory at the same three scales using PyTorch's \smalltt{
  max\_memory\_allocated}. Each baseline has 10 warm-up iterations, followed by
3--10 measured iterations.

For MoE models, memory usage varies across devices due to uneven expert
workloads after dispatch. We evaluate two cases: (1) \emph{Balanced}, using a
feature provided by Megatron Core to ensure uniform distribution, and (2)
\emph{Imbalanced}, where we inject precomputed skewed logits via our MoE Mock
Router (Appendix~\secref{appendix:mock_router}) to create non-uniform
workloads. The MoE Mock Router metrics are \smalltt{br\_min} = 0.71,
\smalltt{br\_max} = 2.16, \smalltt{br\_avg} = 1.48, \smalltt{br\_std} = 0.37,
\smalltt{br\_med} = 1.38, and \smalltt{br\_skew} = 0.90.

Figure~\ref{fig:experiment_e2e_memory} shows that \sysname estimates peak
memory with errors consistently below 0.01\%. Furthermore, \sysname
successfully reproduces OOM errors encountered in the baseline experiments.
\sysname achieves high-fidelity memory estimation by accounting for consumption
across the entire software stack, ensuring that its peak memory predictions are
both comprehensive and accurate.

\noindentpar{End-to-end iteration time with outlier injection.} As
  described in \secref{subsec:graph_calibration}, \sysname supports injecting
  runtime outliers on top of homogeneity-aware timing reuse to emulate
  performance anomalies. We evaluate the accuracy of end-to-end iteration time
  under outlier injection across three model configurations at 1,024- and
  2,048-GPU cluster scales: 235B (with strategy S.B), 503B (with strategy S.A),
  and 1.01T (with strategy S.D).

To establish the ground truth, we first include physical nodes
  flagged for performance degradation in the full-scale cluster runs. After
  warm-up iterations, we profile the execution durations of all operations across
  all ranks over 10 iterations. We then feed the resulting empirical duration
  distributions into \sysname during calibration alongside homogeneity-aware
  timing reuse, constructing a \graphname that captures full-scale execution
  under straggler effects. Finally, \sysname emulates each workload using the
  identical configurations and training scripts for the same number of
  iterations, and we compare the median framework-reported iteration times
  against the full-scale ground truth.

As shown in Figure~\ref{fig:experiment_e2e_stragglers}, \sysname
  accurately emulates end-to-end iteration times under straggler effects,
  achieving an average error of only 0.65\%. Because \graphname accurately
  encodes the necessary replay information and propagates the injected outlier
  delays across all ranks via its dependency edges, \sysname achieves high
  emulation accuracy even under straggler effects.

\subsection{System Efficiency}
\label{subsec:evaluation_efficiency}

\begin{figure}[tb]
  \centering
  \begin{minipage}[t]{0.44\linewidth}
    \centering
    \captionsetup{type=figure}
    \includegraphics[width=\linewidth]{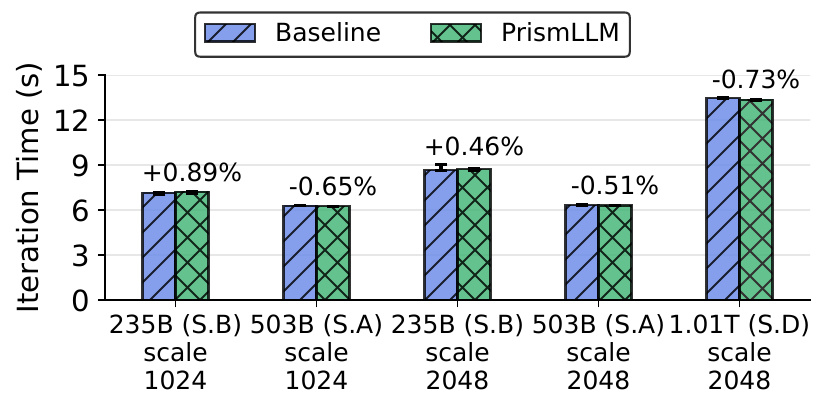}
    \caption{End-to-end iteration time estimation under straggler effects.}
    \Description{Chart comparing median iteration times from full-scale runs with stragglers against estimates from \sysname with outlier injection. The comparison covers the 235B (S.B), 503B (S.A), and 1.01T (S.D) model configurations at cluster scales of 1,024 and 2,048 GPUs. The estimates closely match the measured times.}
    \label{fig:experiment_e2e_stragglers}
  \end{minipage}
  \hfill
  \begin{minipage}[t]{0.54\linewidth}
    \centering
    \captionsetup{type=figure}
    \includegraphics[height=0.4\linewidth]{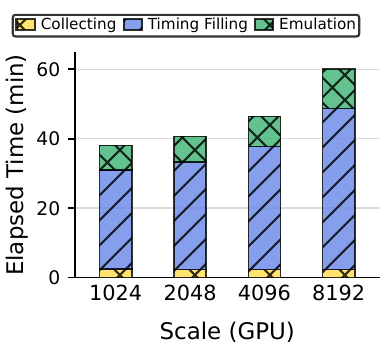}
    \includegraphics[height=0.4\linewidth]{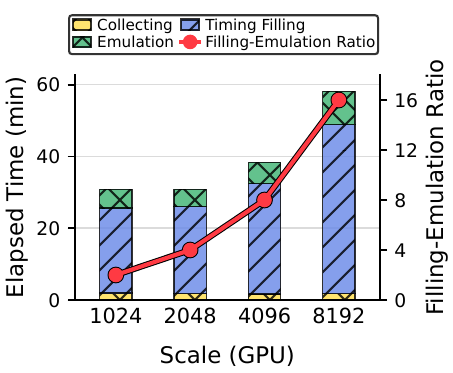}
    \caption{Execution time breakdown for DP scaling (left) and PP scaling (right).}
    \Description{Two charts show \sysname's execution time breakdown as the target scale increases from 1,024 to 8,192 GPUs: DP scaling on the left and PP scaling on the right. Each bar comprises bare graph collection, timing filling and calibration, and hybrid emulation, with timing filling dominating the runtime in both panels. The red line in the PP panel shows the ratio of node pairs used for timing filling to those used for emulation increasing.}
    \label{fig:time_breakdown}
  \end{minipage}
\end{figure}

We evaluate \sysname's physical resource requirements and end-to-end execution
time across target cluster scales (Figure~\ref{fig:time_breakdown}).
Following the common practice of scaling either data parallelism (DP)
  or pipeline parallelism (PP) to accommodate larger workloads, we scale our 503B
  (M.2) model along two dimensions: (1)~along the DP dimension, we scale the
  target cluster from 1,024 to 8,192 GPUs by increasing the number of DP replicas
  while holding the PP degree constant; and (2)~along the PP dimension, we scale
from 1,024 to 8,192 GPUs by proportionally increasing the PP degree from 4 to
64---the deepest pipeline the model supports---while keeping the number of DP
replicas fixed. Across all scales, we measure the end-to-end time
  breakdown covering 24-iteration bare graph collection, timing filling and
  calibration, and final hybrid emulation over the same number of iterations.
  With all proposed optimizations enabled, \sysname keeps the end-to-end
  execution time highly stable as the target scale grows.

For DP scaling, \sysname completes the entire workflow from 1,024 to
  8,192 target GPUs using only a single sandbox node and a single assistant
  node---utilizing <0.2\% of the target GPU resources and achieving >99\%
  resource savings. Bare graph collection takes only about 3~minutes and remains
  constant across scales, as homogeneity-aware graph reuse requires capturing
  only a single DP replica. Similarly, both timing filling and hybrid emulation
  exhibit sublinear time growth: while the cluster scale expands by $16\times$,
  the total time increases by only $1.6\times$. Specifically, homogeneity-aware
  timing reuse keeps the number of profiled slices constant across DP scales. The
  time increase primarily stems from the offline preparation pass that selects
  active virtual ranks in advance (\secref{subsec:virtualize_init}), which grows
  by $8.7\times$ as local shared-memory threads spend more time executing NCCL's
  graph-search routines over larger logical topologies. Computing the participant
  plans across all required process groups for one run at the 8,192-GPU scale
  takes under 4.5~minutes, an acceptable overhead given the substantial
  reductions in initialization time and memory usage relative to emulation
  without virtual rank initialization pruning.

For PP scaling, because pipeline stages are heterogeneous, every
  stage requires timing filling. \sysname can still execute the entire workflow
  from 1,024 ($\text{PP}=4$) to 8,192 ($\text{PP}=64$) target GPUs using just a
  single sandbox--assistant node pair. We use the \textit{filling-emulation node
    ratio} to describe the number of sandbox--assistant node pairs used for timing
  filling relative to that used for final emulation. Using only the single node
  pair required by final emulation (i.e., a 1:1 filling-emulation node ratio)
  forces \sysname to profile all PP stages sequentially: whereas a 512-GPU target
  ($\text{PP}=4$) takes 4 sequential timing filling steps, an 8,192-GPU target
  ($\text{PP}=64$) must run all 64 steps one by one. This increases total runtime
  by about $13\times$---this is how scaling breaks down: as linear time inflation
  rather than memory or compute saturation. To avoid this linear time overhead as
  PP stages grow, \sysname increases the filling-emulation node ratio
  proportionally with the target cluster size (red line in
  Figure~\ref{fig:time_breakdown}). Because proportionally scaled node pairs
  execute multiple slices concurrently and keep the number of sequential
  profiling rounds constant, \sysname keeps the end-to-end workflow time stable
  up to 8,192 GPUs, increasing by only $1.9\times$ over a $16\times$ scale
  expansion, comparable to that of DP scaling.

In summary, the optimizations proposed across both the record and
  replay phases of \sysname---namely, homogeneity-aware graph and timing reuse,
  virtual rank initialization pruning for \emph{all} process groups, and
  collective communication pruning---effectively mitigate scaling overheads,
  thereby ensuring high end-to-end system efficiency.

\subsection{Micro benchmarks}
\label{subsec:evaluation_micro}

\begin{figure}[tb]
  \centering
  \begin{subfigure}[t]{0.69\linewidth}
    \centering
    \includegraphics[width=\linewidth]{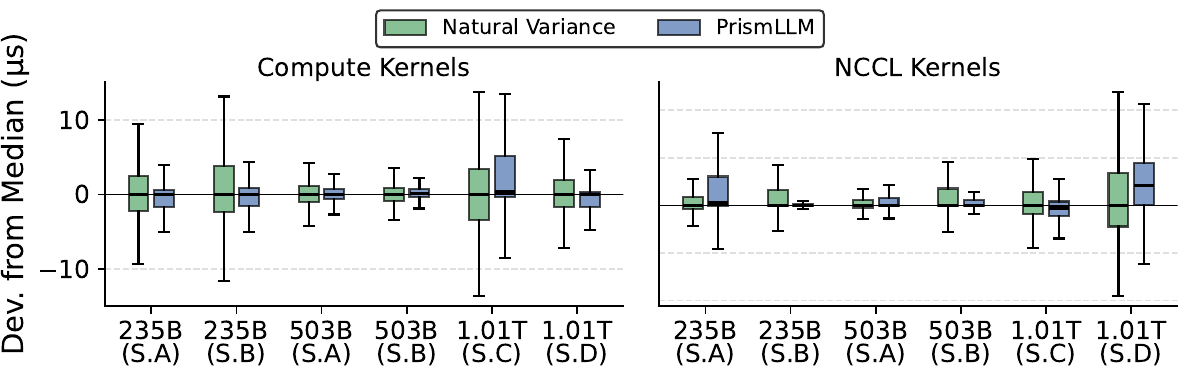}
    \caption{Kernel duration deviations from baseline medians.}
    \label{subfig:graph_accuracy_kernel}
  \end{subfigure}
  \hfill
  \begin{subfigure}[t]{0.29\linewidth}
    \centering
    \includegraphics[width=\linewidth]{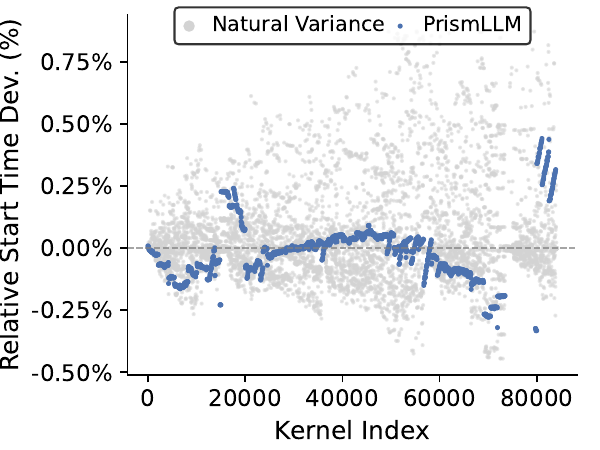}
    \caption{Kernel start-time deviations.}
    \label{subfig:graph_accuracy_start}
  \end{subfigure}
  \caption{Kernel-level timing fidelity of \sysname at 1,024-rank scale.}
  \Description{Two plots compare \sysname with natural hardware variance at 1,024-rank scale. The left plot shows compute and NCCL kernel duration deviations from baseline medians in microseconds across model and parallelization configurations. The right plot shows kernel start-time deviations from baseline medians as percentages of total iteration time for the 503B model with strategy S.B.}
  \label{fig:execution_fidelity}
\end{figure}

\noindentpar{Execution fidelity.} To validate the execution accuracy inside
\sysname, we compare its PyTorch Profiler traces against training baselines. We
perform a fine-grained, kernel-level comparison using the absolute deviation in
kernel duration. For the training baseline, we sample 32 ranks within the same
DP group and calculate the median execution duration of each kernel. The
inherent hardware jitter---the deviation of actual execution times from this
median---is represented by the \textit{Natural Variance} boxes. Then, we
compare the emulated kernel durations against the corresponding baseline
medians, with the resulting errors shown by the \textit{\sysname} boxes. As
shown in Figure~\ref{subfig:graph_accuracy_kernel}, for all configurations at
the scale of 1024 ranks, the emulated variance falls within the range of
natural hardware variance. Beyond duration, we evaluate scheduling fidelity via
relative kernel start times (normalized by total iteration time).
Figure~\ref{subfig:graph_accuracy_start} visualizes the 1024-rank 503B (S.B)
case, plotting the natural variance across 32 baseline ranks (gray) against
\sysname's deviation from the baseline median (blue). \sysname tightly tracks
the median, bounded within the ±0.5\% natural hardware jitter. Across all
configurations, the maximum observed start time error is 3.64\%. This
kernel-level fidelity underpins the accurate prediction of configuration
changes that alter kernel composition, such as disabling Flash Attention
(Table~\ref{tab:new_feature}).

\begin{figure}[tb]
  \centering
  \begin{subfigure}[t]{0.49\linewidth}
    \centering
    \includegraphics[height=0.4\linewidth]{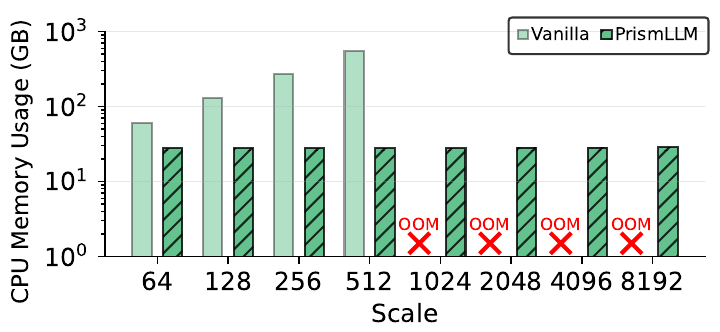}
    \caption{Overall CPU memory overhead.}
    \label{subfig:bootstrap_reduction_cpu}
  \end{subfigure}
  \hfill
  \begin{subfigure}[t]{0.49\linewidth}
    \centering
    \includegraphics[height=0.4\linewidth]{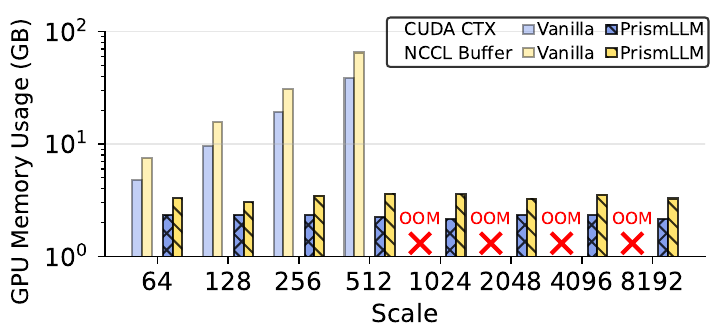}
    \caption{GPU memory overhead per assistant GPU.}
    \label{subfig:bootstrap_reduction_gpu}
  \end{subfigure}
  \caption{Memory usage with virtual rank initialization pruning in \sysname.}
  \Description{Two plots compare the standard NCCL baseline with \sysname at scales from 64 to 8192 logical ranks. They show total CPU memory usage and GPU memory usage per assistant GPU broken down into CUDA contexts and NCCL buffers.}
  \label{fig:bootstrap_reduction_memory}
\end{figure}

\noindentpar{Virtual rank initialization pruning.} To enable efficient
large-scale emulation, \sysname features a customized initialization mechanism.
We evaluate its performance by measuring the time and memory required to
initialize and execute a global barrier operation across varying cluster scales
(64 to 8192 ranks). The testbed consists of two 8-GPU nodes: one acts as the
sandbox node (running one sandbox rank per GPU), while the other serves as the
assistant node and hosts all virtual ranks. We compare \sysname against a
baseline that uses standard NCCL (via a shared \smalltt{NCCL\_HOSTID} value),
where each virtual rank maintains an independent process and CUDA context.

\begin{figure}[tb]
  \centering
  \begin{minipage}[t]{0.48\linewidth}
    \centering
    \captionsetup{type=figure}
    \includegraphics[height=0.5\linewidth]{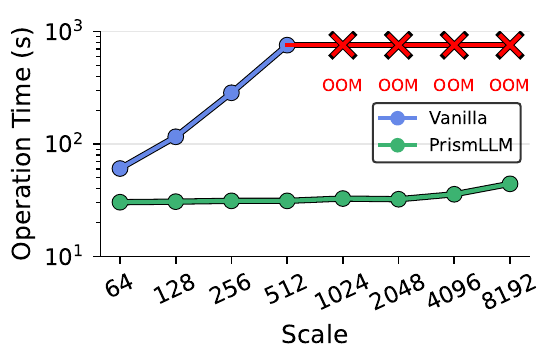}
    \caption{Initialization time with virtual rank initialization pruning in \sysname.}
    \Description{A plot compares the standard NCCL baseline with \sysname at scales from 64 to 8192 logical ranks, showing the time to initialize and complete one global barrier.}
    \label{fig:bootstrap_reduction_time}
  \end{minipage}
  \hfill
  \begin{minipage}[t]{0.48\linewidth}
    \centering
    \captionsetup{type=figure}
    \includegraphics[height=0.5\linewidth]{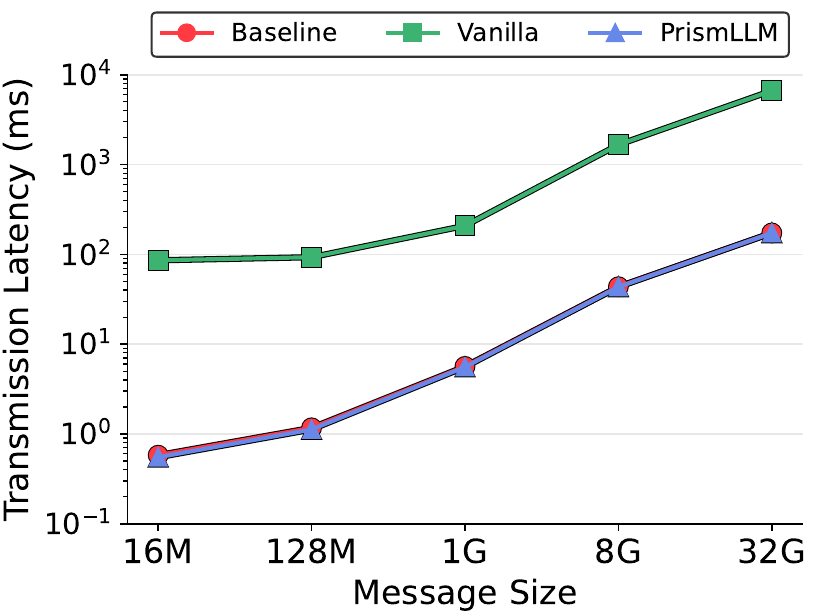}
    \caption{Effect of collective communication pruning on allreduce transmission latency at 32-rank scale.}
    \Description{Line chart comparing allreduce transmission latency at 32 ranks for message sizes from 16 MB to 32 GB. \sysname closely follows the physical baseline, while Vanilla emulation has substantially higher latency.}
    \label{fig:transmission_reduction}
  \end{minipage}
\end{figure}

As shown in Figures~\ref{fig:bootstrap_reduction_memory}
and~\ref{fig:bootstrap_reduction_time}, the baseline incurs linearly scaling
overheads. At merely 512 ranks, initializing CUDA contexts, establishing
communication channels, and allocating NCCL buffers consume 544.5~GB of CPU
memory and 103.7~GB of GPU memory. Consequently, initialization takes 756.0
seconds to complete, and the baseline inevitably encounters OOM errors at
scales $\ge$ 1024 ranks. In contrast, \sysname maintains a strictly
\textit{constant} resource footprint regardless of the logical cluster size. By
strategically allocating resources only for neighboring virtual ranks (i.e.,
direct algorithmic neighbors of sandbox ranks), \sysname eliminates large
numbers of redundant CUDA contexts and NCCL buffers. Consequently, \sysname
successfully initializes an 8192-rank emulation in just 44.3 seconds, occupying
a mere 5.4~GB of GPU memory per physical assistant GPU.

\noindentpar{Collective communication pruning.} While the initialization
pruning above minimizes \textit{initialization} overhead, \sysname must also
mitigate \textit{runtime} resource contention during collective operations.
Using the same two-node testbed, we evaluate allreduce execution. In Vanilla
emulation, virtual ranks blindly replay standard NCCL kernels, intensely
contending with sandbox ranks for streaming multiprocessors (SMs) and
memory/PCIe bandwidth, thereby skewing execution timing. To ensure high
fidelity, \sysname prunes redundant data transmissions among virtual ranks,
preserving only cross-node RDMA traffic.

Figure~\ref{fig:transmission_reduction} shows the transmission latency across
varying message sizes at a 32-rank scale, while
Table~\ref{tab:allreduce_latency} presents the full results across message
sizes from 16~MB to 32~GB and emulation scales up to 128 ranks. Without
pruning, Vanilla emulation suffers a latency inflation of up to $148\times$ at
32 ranks. By applying collective communication pruning, \sysname closely tracks
the physical baseline, yielding a maximum error of 0.2\% against real
executions at this scale.

\begin{table}[tb]
  \centering
  \scriptsize
  \caption{Allreduce transmission latency (ms) for the physical baseline, Vanilla emulation, and \sysname.}

  \label{tab:allreduce_latency}
  \begin{tabular}{@{} l rrrr rrrr rrrr @{}}
    \toprule
    \multirow{2}{*}{\textbf{Size}}                 &
    \multicolumn{4}{c}{\textbf{Baseline}}          &
    \multicolumn{4}{c}{\textbf{Vanilla Emulation}} &
    \multicolumn{4}{c}{\textbf{\sysname}}                                                                                                                                                \\
    \cmidrule(lr){2-5} \cmidrule(lr){6-9} \cmidrule(lr){10-13}
                                                   & \textbf{16} & \textbf{32} & \textbf{64} & \textbf{128}
                                                   & \textbf{16} & \textbf{32} & \textbf{64} & \textbf{128}
                                                   & \textbf{16} & \textbf{32} & \textbf{64} & \textbf{128}                                                                              \\
    \midrule
    16M                                            & 0.27        & 0.58        & 0.41        & 0.50         & 0.28   & 85.92   & 181.98   & 639.44   & 0.27   & 0.55   & 0.27   & 0.97   \\
    128M                                           & 1.22        & 1.17        & 1.56        & 2.17         & 1.22   & 92.98   & 734.74   & 981.97   & 1.18   & 1.12   & 1.82   & 3.97   \\
    1G                                             & 6.32        & 5.65        & 6.40        & 8.11         & 6.38   & 208.17  & 1070.08  & 4351.50  & 6.19   & 5.56   & 6.34   & 12.72  \\
    8G                                             & 47.72       & 43.61       & 44.34       & 45.04        & 47.92  & 1663.51 & 4256.79  & 12742.20 & 47.21  & 43.34  & 44.71  & 67.04  \\
    32G                                            & 189.09      & 173.25      & 177.08      & 177.62       & 189.22 & 6655.51 & 16950.36 & 50815.80 & 187.96 & 173.11 & 176.70 & 267.42 \\
    \bottomrule
  \end{tabular}
  \begin{minipage}{\textwidth}
    \vspace{0.5em}
    \centering
    \textit{*Columns within each method report results at 16, 32, 64, and 128 ranks.}
  \end{minipage}
\end{table}

The results highlight the severe resource contention inherent in the Vanilla
emulation approach, which suffers from exponential latency inflation as the
message size and cluster scale grow. In contrast, \sysname's collective
communication pruning strategy consistently minimizes overhead, closely
tracking the physical baseline across all tested configurations. However, as
the scale reaches 128 ranks, \sysname exhibits noticeable latency deviations
for large messages (e.g., 267.42~ms vs. 177.62~ms at 32~GB). This is expected:
hosting all virtual ranks on a single assistant node saturates its PCIe
bandwidth and SMs. As detailed in Section~\ref{subsec:workflow}, \sysname
gracefully resolves this hardware bottleneck by scaling out virtual ranks
across multiple assistant nodes to restore emulation fidelity.

\noindentpar{Inter-slice timing calibration.} To demonstrate the effectiveness
of inter-slice timing calibration, we chose the 235B model with strategy S.B
and replayed the \graphname before and after inter-slice timing calibration.
The emulated iteration time before inter-slice timing calibration quickly
dropped from 5.7s to 5.13s, demonstrating that without proper alignment, the
emulation error can easily go higher than 10\%.

\subsection{Comparison with Simulators}
\label{subsec:evaluation_simulators}

\begin{figure}[tb]
  \centering
  \includegraphics[width=0.6\linewidth]{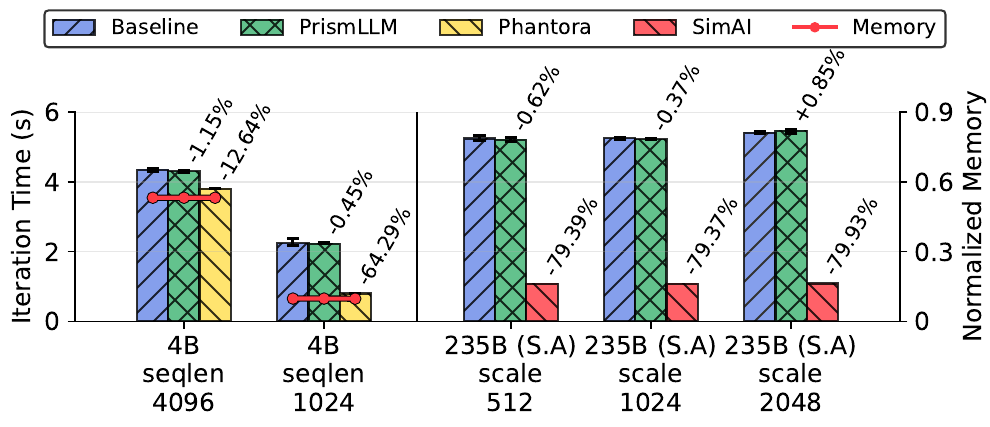}
  \caption{Comparison of \sysname with Phantora and SimAI.}
  \Description{Grouped bars compare iteration time in seconds against measured baselines. The left two groups compare \sysname and Phantora for the 4B model at sequence lengths of 4096 and 1024. The right three groups compare \sysname and SimAI for the 235B model with strategy S.A at 512, 1024, and 2048 ranks, without memory results. \sysname closely matches measured iteration times, while both simulators underestimate them.}
  \label{fig:end_to_end_simulators}
\end{figure}

\noindentpar{Phantora.} We evaluate \sysname against Phantora on small-scale
tasks within its supported configurations. We assess their accuracy in
estimating iteration time and peak memory allocation under identical model and
parallelization configurations. The setup is a 32-GPU testbed using the Qwen 3
4B configuration, with the TP size set to 4.

As shown in Figure~\ref{fig:end_to_end_simulators}, \sysname outperforms
Phantora in both metrics. Notably, while Phantora shows a 12\% error in typical
cases, its prediction fidelity drops significantly as the sequence length
decreases—with the error rising to 64\% as the GPU-side computational pressure
is decreased. This discrepancy primarily stems from two factors: (1) Phantora
approximates actual kernel duration using raw network bandwidth, failing to
account for the complex, multi-step nature of collective kernels; and (2)
Phantora launches multiple mock processes on CPU to represent individual
training ranks, advancing the simulation via inter-process communication with a
central simulator. Due to the CPU contention among multiple processes, Phantora
cannot accurately capture the impact of CPU-side Python interpreter latency on
the overall execution duration. As a result, Phantora defaults to omitting
these latency components, leading to a systematic underestimation of the total
iteration time.

\noindentpar{SimAI.} To further evaluate the performance of simulators in
large-scale MoE training scenarios, we feed the model configurations and
parallelization strategies into SimAI~\cite{simai} and compare its results
against \sysname. Notably, SimAI does not support memory allocation simulation
and is therefore excluded from the memory-related comparison.

Figure~\ref{fig:end_to_end_simulators} presents the results for the 235B (M.1)
model with strategy S.A at different scales; comprehensive results for other
configurations are deferred to Appendix~\secref{app:simai_all}. SimAI deviates
substantially from the baseline, with an average error of 77.2\%. We attribute
this discrepancy to two primary factors. First, while SimAI relies on workload
files to describe the training process, the workload-file design fails to
capture the inter-rank dependencies and asymmetric behaviors inherent in
pipeline model parallelism (PP), omitting the communication bubbles associated
with PP. Second, SimAI neglects the unique characteristics of MoE models
compared to dense ones, ignoring the computational overheads of MoE operations
such as gating, permute, dispatch, and combine. These findings underscore that
simulators often require manual effort to maintain and update simulation models
for evolving architectures.

\section{Use Cases and Experience}
\label{sec:usecases}

Our production use cases build on two properties of \sysname. First, it
faithfully emulates the \emph{normal} (average) behavior of a training job at
scale; this serves both as the metric for comparing configurations and as the
expected baseline against which anomalies stand out. Second, the environment is
isolated from production and repeatable, so heavyweight inspection that
production prohibits can be applied freely. The cases below specialize these
two properties.

\noindentpar{Tuning training configurations.} Maximizing the
  utilization of device memory capacity is essential for achieving high training
  efficiency. This efficiency is commonly quantified by Model FLOPs Utilization
  (MFU). Two classes of strategies significantly shape memory usage:
  \emph{distributed strategies}, which determine how the model is partitioned
  across GPUs, and \emph{memory strategies}, which orchestrate trade-offs across
  compute, interconnects, and host resources via techniques like selective
  recomputation and offloading. Yet, finding optimal strategies for a specific
  workload and hardware platform is non-trivial due to the massive configuration
  space. For our production workloads, jointly searching both dimensions yields
  over $10^5$ raw combinations, of which more than a thousand are feasible.
  Evaluating these candidates at full scale is prohibitively expensive, and
  evolving workload characteristics across training stages make this tuning a
  recurring burden.

\begin{table}[tb]
  \centering
  \footnotesize
  \caption{Emulation accuracy under different training configurations.}
  \label{tab:new_feature}
  \begin{tabular}{@{} l rrrr @{}}
    \toprule
    \multirow{2}{*}{\textbf{Optimization}}          &
    \multicolumn{2}{c}{\textbf{Iteration Time (s)}} &
    \multicolumn{2}{c}{\textbf{Peak Memory (GB)}}                                                                                         \\
    \cmidrule(lr){2-3} \cmidrule(lr){4-5}
                                                    & \textbf{Baseline}       & \textbf{\sysname} & \textbf{Baseline} & \textbf{\sysname} \\
    \midrule
    Baseline                                        & 5.62 ($+0.29$/$-0.08$)  & 5.70              & 133.76            & 133.76            \\
    Flash Attention Off                             & 7.74 ($+0.24$/$-0.08$)  & 7.70              & 135.22            & 135.22            \\
    P2P Overlap Off                                 & 5.71 ($+0.38$/$-0.09$)  & 5.82              & 133.74            & 133.74            \\
    Offload Optimizer                               & 11.98 ($+0.51$/$-0.24$) & 11.61             & 112.07            & 112.07            \\
    Recompute                                       & 7.07 ($+0.37$/$-0.08$)  & 7.21              & 98.13             & 98.13             \\
    \bottomrule
  \end{tabular}
\end{table}

\begin{figure}[tb]
  \centering
  \includegraphics[height=0.39\linewidth]{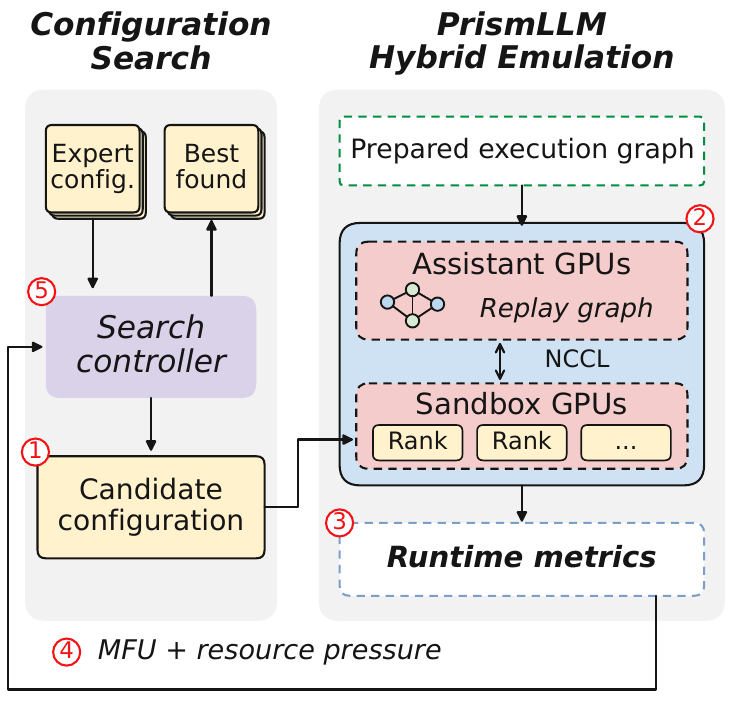}
  \hfill
  \includegraphics[height=0.39\linewidth]{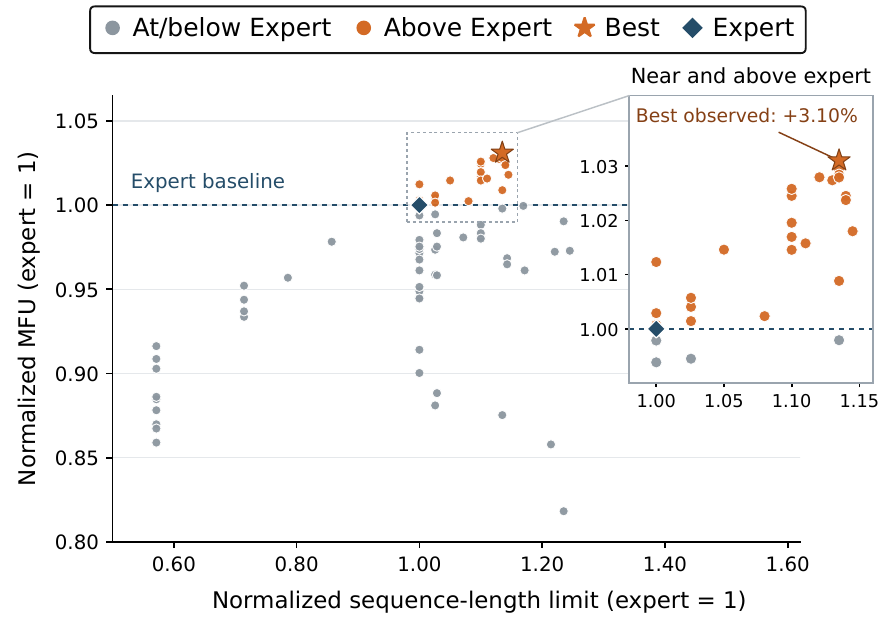}
  \caption{Resource-guided training configuration search with \sysname (left) and the normalized training configuration landscape (right).}
  \Description{The left diagram shows a search controller proposing candidate configurations and hybrid emulation evaluating them with a prepared graph. Runtime metrics feed MFU and resource pressure back to the controller. The right plot shows normalized MFU against normalized sequence-length limits relative to an expert baseline, with an inset highlighting configurations near and above that baseline.}
  \label{fig:configuration_search_workflow}
\end{figure}

To alleviate this burden, \sysname supports iterative configuration
  tuning with a small GPU footprint while preserving production-scale behavior.
  In our deployment, this reduces evaluation GPU-hours by 87.5\% compared with
  downscaled trials running a single model replica. The small footprint also
  enables tuning trials to use idle GPU allocations, improving overall
  utilization of the test cluster's GPU resources. Table~\ref{tab:new_feature}
  shows that \sysname accurately reproduces iteration time and peak GPU memory
  usage across different training optimizations.
  Figure~\ref{fig:configuration_search_workflow} (left) shows the evaluation
  loop. Starting from an initial configuration, a search controller proposes a
  candidate's parallelism, selective recomputation, offloading, and
  sequence-length settings \circled{1}. The sandbox executes the candidate's
  training code, while assistant GPUs replay the other ranks' execution graphs
  and participate in communication \circled{2}. The emulation reports MFU and
  runtime measurements \circled{3}, including SM and tensor core activity,
  allocated and reserved GPU memory, traffic over PCIe, NVLink, and the network,
  host resource use, execution timing and imbalance along parallelism dimensions.
  The available resource measurements are normalized against resource capacities
  to form a resource-pressure vector $\boldsymbol{\phi}$ \circled{4}, which,
  together with MFU, is fed back to the controller to guide the next proposal
  \circled{5}.

This resource-aware feedback loop enables the controller to prune
  the search space and make targeted adjustments rather than searching blindly.
  Within the controller, MFU evaluates overall candidate quality, whereas the
  resource-pressure vector $\boldsymbol{\phi}$ diagnoses current bottlenecks and
  reveals remaining resource headroom. The controller combines these observations
  with a memory model and the expected resource effects of configuration changes
  to prioritize feasible candidates. After each trial, the measured resource
  changes continuously update the model and directly drive the next concrete
  proposal, rather than merely explaining performance post hoc.

Figure~\ref{fig:configuration_search_workflow} (right) summarizes
  the evaluated configurations, normalizing each candidate's MFU and
  sequence-length limit to its corresponding expert-tuned baseline at $(1,1)$.
  The search improved on the baseline, with a best observed relative MFU gain of
  3.10\%. At the scale of published cloud compute cost estimates for frontier
  models~\cite{stanford_ai_index_2025}, an MFU gain of this magnitude could
  translate into millions of dollars in training compute savings. Finding this
  configuration illustrates how \sysname's runtime measurements guide tuning by
  exposing differences that peak allocated memory alone obscures. The search
  initially enabled expert-parallel dispatch caching without changing the
  buffer-release policy. MFU fell despite an almost unchanged peak allocation.
  This regression prompted the controller to keep caching enabled and compare
  buffer-release policies at a fixed sequence-length limit. The two policies had
  the same peak allocated memory, yet one achieved higher MFU with lower reserved
  memory and more physical GPU memory headroom. The controller used that headroom
  to test a larger sequence-length limit with the same cache and synchronization
  settings, yielding the highest observed MFU without full-layer recomputation or
  activation offloading.

\noindentpar{Memory analysis.} Beyond configuration tuning,
  \sysname provides an isolated, repeatable sandbox to investigate underlying
  memory allocator dynamics via on-demand memory snapshots. We applied this
  capability to three real-world scenarios: estimating memory requirements,
  diagnosing fragmentation, and identifying memory bugs.

\emph{MoE memory estimation}: token routing is dynamic and shifts across
training stages; early-stage imbalance can lengthen iterations by over 5\% and
swing memory by up to 20~GB, so aggressive memory configurations risk OOM while
conservative ones waste GPUs. \sysname allows engineers to plug in a mock
router that replays production-observed imbalance features
(Appendix~\secref{appendix:mock_router}) and estimates the dynamic memory
accurately (Figure~\ref{fig:experiment_e2e_memory}).

\emph{Fragmentation attribution}: after adopting a new attention kernel, the
maximum trainable sequence length unexpectedly dropped by 13\%; snapshots
showed that activations became fine-grained, and small tensors occupied
segments that previously held large contiguous activations, lowering reuse and
raising the reserved watermark by 9\% of the per-rank GPU memory.

\emph{Memory bugs}: in one code version, temporary tensors were not released in
time and the reserved-minus-allocated gap grew abnormally; snapshots pinpointed
the leak, and the fix shrank the gap from 19\% to 8\% of the per-rank GPU
memory.

\noindentpar{Targeted cluster health checks.} Gray failures are notoriously
hard to pinpoint: benchmark suites~\cite{superbench} hardly cover the full
software stack and its corner cases. In one incident, a job launched with an
iteration time of 6.38s, against an expected median of 5.63s drawn from
historical monitoring (a 14\% slowdown). The infrastructure team collected the
\graphname offline as the normal behavior, then screened the cluster with
\sysname before relaunching the job: machines were paired, one node as sandbox
running the real training code and the other as assistant, swapping roles
afterwards---each pass costing about 20 minutes.

One sandbox node reported 6.40s versus a 5.70s median across the other pairs;
inspection revealed thermal throttling that down-clocked its GPUs to 900MHz.
The job recovered after the node was taken offline. Generic suites miss this
issue because it manifests only under the target job's sustained execution mix.
This incident illustrates how deviations from the reproduced
  baseline can support a health check for a specific training workload. Extending
  this approach to other types of cluster faults remains a direction for future
  work.

\noindentpar{Fine-grained profiling and tool development.} Heavyweight or
immature observability tools are often prohibited online due to their overhead,
pressure on the logging system, or risk of their own bugs. \sysname's isolated
environment lifts these constraints: all results in this paper were collected
with unmodified \smalltt{nvidia-smi}, PyTorch Profiler, and NVIDIA Nsight
Systems. The same environment also supports building new tools: in one case, we
developed tracing tools on \sysname. Debugging concentrated on injection: CUDA
version incompatibilities, and the tuning of hook points and trigger
conditions. Downscaled runs could reproduce some of these issues, but the hooks
fire only on code paths exercised under the production configuration (e.g.,
specific sequence-length ranges), so the most stable way to iterate was against
the unmodified configuration---which \sysname provides on a few GPUs.

\section{Related Work}
\label{sec:related}

\noindentpar{Answering pretraining questions with fewer GPUs.} Planning,
tuning, and debugging large-scale training all call for tools that consume far
fewer GPUs than the target cluster, but they tolerate different levels of
abstraction. Simulators answer questions from a \textit{description} of the
system: analytical ones~\cite{astrasim1,astrasim2} need only operator counts,
bandwidths, and topology---usable before any hardware exists---while
profiling-driven ones~\cite{vtrain,distsim,simai,daydream,dpro,multiverse}
replay kernel timings measured on real GPUs inside a model. At near-zero
per-configuration cost, they are the right tool for sweeping design spaces and
ranking candidates.

A method can answer only what its input encodes: allocator fragmentation, SM
scheduling, and interpreter latency have no counterparts in a description;
modeling them faithfully amounts to reimplementing the system
(\secref{subsec:simulating}). Reproducing these behaviors requires the
\textit{artifact} itself---real code on real hardware. Downscaled observations
may not transfer back because the configuration changes
(\secref{subsec:downscaling}); full-scale runs require the target GPU count.
\sysname preserves the runnable code and target hardware model, giving up only
quantity---the physical concurrency of all ranks: enough to determine whether
this stack at scale meets its iteration-time target, fits in memory, and
reproduces production failures.

\noindentpar{Emulating scale in other domains.} Emulating a large system on few
machines has a long history; each lineage gives up quantity through a different
mechanism. \emph{Time dilation}~\cite{timedilation,diecast} slows every node's
clock so that the same hardware appears to offer $N\times$ the bandwidth or
compute---in that setting, scale is a rate that a clock can fake. In
pretraining, scale is not a rate: a larger job changes the parallelism
configuration and every rank's execution graph, which no rescaling of kernel
durations reproduces.

\emph{Network emulation}~\cite{emulab,modelnet,kollaps} multiplexes many
lightweight endpoints per machine and synthesizes the fabric between them---the
endpoints still execute the real application. A faithfully executing rank
cannot be made cheap this way: multiplexed onto one GPU, its kernels no longer
run at native speed. \sysname instead stops executing the other ranks and
replays them, which works because DP replicas are statistically symmetric and
the communication schedule is static; datacenter traffic, emerging from all
flows jointly, offers no such representative subset.

\emph{Boundary emulation}: CrystalNet~\cite{crystalnet} boots real switch
firmware within a change's scope and mocks the rest---the principle behind
\sysname's neighbor-rank pruning (\secref{subsec:virtualize_comm});
FireSim~\cite{firesim} and SimBricks~\cite{simbricks} push this to cycle level.
But each mocked device acts from its own static snapshot, so the surroundings
scale by replication; a virtual rank's actions---what to send, and
when---follow from the dependency structure of the entire job at target scale,
which is why \sysname must first capture the execution graph (Phase~1).

\emph{Learned approximation}: MimicNet~\cite{mimicnet} simulates one cluster at
full fidelity and approximates the rest with models trained on it; \sysname
replaces the approximation with execution---virtual ranks send NCCL traffic at
full volume---so a refresh is a minute-level re-collection, not retraining.

\emph{Record-and-replay}~\cite{liblog,friday,r2} re-injects recorded inputs to
reproduce causality at arbitrary speed; \sysname likewise re-injects
communication tensors and builds on GPU
checkpointing~\cite{criu,cuda_checkpoint}, but timing itself is the observable
here, so durations are replayed exactly and content is recorded only at the
sandbox boundary.

\noindentpar{Emulating LLM training.} The closest systems differ in how much of
the real artifact survives into the loop. Phantora~\cite{phantora} runs the
unmodified program on mock CUDA and NCCL backends; execution stops being real
at the backend, which explains its growing error as GPU-side pressure drops
(\secref{subsec:evaluation_simulators}). Echo~\cite{echo} collects traces on a
single device much as Phase~1 does, but replays them into a simulator with
modeled communication. Chakra~\cite{chakra} and Mystique~\cite{mystique} replay
traces as standalone benchmarks rather than sustaining full-scale interaction
around real ranks. \sysname keeps the real program, real GPUs, and real NCCL
end to end, replacing only remote ranks' computation with measured durations.

\section{Conclusion}
\label{sec:conclusion}

We present \sysname, a system that faithfully emulates large-scale LLM training
using a small fraction of its hardware. By capturing a high-fidelity execution
graph on a few GPUs and replaying it around the ranks of interest, \sysname
emulates deployment-scale performance and memory behavior while running
unmodified training code. In our deployment, engineers use \sysname
  to evaluate training changes, diagnose runtime issues, and develop
  observability tools. We are extending \sysname toward broader training
frameworks and more asymmetric workloads and we\ifAnonymous{ will }{ }release
the system to support future research and industry adoption.

\begin{acks}

  We thank our reviewers for their insightful comments and suggestions, which
  significantly improved this paper. We sincerely thank the Wan pretraining
  infrastructure team for their invaluable feedback on \sysname.

\end{acks}

\bibliographystyle{ACM-Reference-Format}
\bibliography{reference}

\appendix

\section{Coordinator and Priority-Based Context Switching Algorithm}
\label{app:priority-switching}

The coordinator employs a greedy scheduling strategy that maximizes system-wide
progress by prioritizing processes with the highest potential for unblocking
other waiting operations. Algorithm~\ref{alg:gpu-switching} presents the core
switching logic.

When a process encounters a collective operation that cannot be executed
immediately, the coordinator must decide which frozen process to activate on
the same GPU. The selection process (lines 3--22) applies several filtering
criteria to identify viable candidates. First, it excludes finished processes
and those pinned to different GPUs (lines 7--12), ensuring only relevant
processes compete for the target GPU. Most critically, it filters out processes
whose head-of-line blocking operation is not ready (lines 13--15), as switching
to such processes would yield no forward progress.

Among eligible candidates, the coordinator selects the process with the maximum
number of pending operations (lines 16--19). This greedy heuristic is based on
the insight that processes with more pending work are likely to unblock a
larger number of dependent operations when given GPU time, thereby maximizing
overall system throughput.

The main coordination loop (lines 23--36) handles incoming collective
notifications by first updating the pending operation counts for all waiting
ranks (line 25). If all participants in a collective operation are currently
resident on GPUs, the operation executes directly without context switching
(lines 26--27). Otherwise, the coordinator performs context switching by
freezing the requesting process, activating the selected candidate, and storing
communication tensors to disk (lines 31--33) for later retrieval when all
participants become available.

This approach effectively transforms the limited GPU resources into a larger
virtual execution environment, enabling trace generation for large-scale
training scenarios using significantly fewer physical GPUs while maintaining
correctness through careful dependency tracking and progress-oriented
scheduling.

\begin{algorithm}[htbp]
\small
\caption{Priority-Based GPU Context Switching}
\label{alg:gpu-switching}
\begin{algorithmic}[1]
\STATE \textbf{Input:} Trigger rank $r_t$, Worker status $S$
\STATE \textbf{Output:} Best switching candidate

\STATE \textbf{function} \textsc{SelectSwitch}($r_t$)
    \STATE $gpu \leftarrow S[r_t].gpu\_id$
    \STATE $max\_pending \leftarrow 0$, $best \leftarrow \textsc{NULL}$
    
    \FOR{$rank \in S$}
        \IF{$S[rank].status = \textsc{FINISHED}$}
            \STATE \textbf{continue}
        \ENDIF
        \IF{$S[rank].gpu\_id \neq gpu$}
            \STATE \textbf{continue}
        \ENDIF
        \IF{$S[rank].head\_op.status \neq \textsc{READY}$}
            \STATE \textbf{continue}
        \ENDIF
        
        \IF{$S[rank].pending\_ops > max\_pending$}
            \STATE $max\_pending \leftarrow S[rank].pending\_ops$
            \STATE $best \leftarrow rank$
        \ENDIF
    \ENDFOR
    \STATE \textbf{return} $best$
\STATE \textbf{end function}

\STATE \textbf{function} \textsc{HandleCollective}($msg$)
    \STATE $r \leftarrow msg.sender$
    \STATE \textsc{UpdatePendingOps}$(msg.remaining\_ranks)$
    
    \IF{\textsc{AllParticipantsOnGPU}$(msg.participants)$}
        \STATE \textsc{ExecuteDirect}$(msg)$
    \ELSE
        \STATE $candidate \leftarrow \textsc{SelectSwitch}(r)$
        \IF{$candidate \neq \textsc{NULL}$}
            \STATE \textsc{FreezeRank}$(r)$
            \STATE \textsc{ActivateRank}$(candidate)$
            \STATE \textsc{StoreTensors}$(msg.tensors)$
        \ENDIF
    \ENDIF
\STATE \textbf{end function}

\STATE \textbf{function} \textsc{UpdatePendingOps}($waiting\_ranks$)
    \FOR{$rank \in waiting\_ranks$}
        \STATE $S[rank].pending\_ops \leftarrow S[rank].pending\_ops + 1$
    \ENDFOR
\STATE \textbf{end function}

\end{algorithmic}
\end{algorithm}

\section{MoE Mock Router}
\label{appendix:mock_router}

For MoE models, memory allocation typically varies across devices because
experts on different devices may process uneven amounts of data after the
dispatch phase. The final data distribution is determined by the gating
mechanism. After the input data on each rank passes through the gating layer,
each rank obtains the probabilities (or logits) of its local tokens being
dispatched to each expert. All ranks participating in expert parallelism then
perform communications to exchange these local probabilities. By combining
these probabilities with the routing strategy (e.g., \smalltt{top\_k}), the
final global data distribution is determined.

To control the non-uniform dispatching of MoE models and observe the resulting
memory footprint after dispatch, we designed the MoE Mock Router. This
component utilizes the Balance Ratio (\smalltt{br}) to regulate distribution
probabilities. The \smalltt{br} represents the ratio of the actual data volume
possessed by a specific rank to the volume it would possess under a perfectly
uniform distribution. A $\smalltt{br} > 1$ indicates that the rank is
over-utilized relative to the average. Since each microbatch must pass through
all layers within a single iteration, multiple gating operations occur, each
requiring control via \smalltt{br}. We characterize the distribution of
\smalltt{br} in one iteration using statistical metrics including
\smalltt{br\_min} (minimum), \smalltt{br\_max} (maximum), \smalltt{br\_avg}
(average), \smalltt{br\_std} (standard deviation), \smalltt{br\_med} (median),
and \smalltt{br\_skew} (skewness). Based on these statistics, the MoE Mock
Router derives the \smalltt{br} distribution and pre-calculates the logits for
all ranks within an expert parallel group. These pre-calculated logits are
subsequently injected into the logits tensor generated by the gating mechanism
during each invocation.

To ensure that these pre-calculated values do not incur additional GPU memory
overhead, they are stored in pinned host memory. The Mock Router performs an
asynchronous in-place copy to the GPU-resident logits tensor only when the
gating operation is triggered, effectively overwriting the original values
without allocating new device buffers.

\section{Extended results of \sysname vs. SimAI}
\label{app:simai_all}

\begin{figure}[tb]
  \centering
  \includegraphics[width=\linewidth]{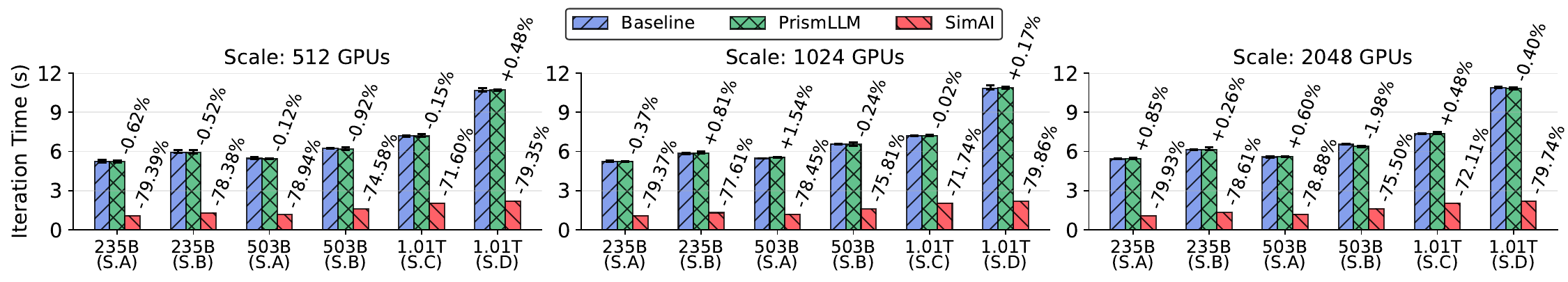}
  \caption{End-to-end iteration time estimates from \sysname and SimAI compared with measured baselines.}
  \Description{Three grouped bar charts compare iteration times in seconds at scales of 512, 1024, and 2048 GPUs. Each panel includes the 235B and 503B models under strategies S.A and S.B, and the 1.01T model under strategies S.C and S.D. \sysname remains close to the baseline across configurations, while SimAI consistently underestimates iteration time.}
  \label{figure:simai_all}
\end{figure}

Figure~\ref{figure:simai_all} presents the comparison between \sysname and
SimAI across all model and strategy configurations used in
Section~\ref{subsec:evaluation_simulators}. SimAI is excluded from the memory
evaluation as it does not support memory allocation simulation. Consistent with
the analysis in Section~\ref{subsec:evaluation_simulators}, SimAI
underestimates iteration time across all configurations, with an average error
of 77.2\%, mainly because its workload files omit the communication bubbles of
PP and the computation overheads of MoE-specific operations.

\end{document}
\endinput